\documentstyle{mn}

\def\Lya{Ly$\alpha\ $}
\def\Lyb{Ly$\beta\ $}

\def\LCDM{$\Lambda$CDM\ }
\def\LCDMns{$\Lambda$CDM}
\def\LCDML{$\Lambda$CDM$_{\rm L}$\ }
\def\LCDMH{$\Lambda$CDM$_{\rm H}$\ }
\def\LCDMLns{$\Lambda$CDM$_{\rm L}$}
\def\LCDMHns{$\Lambda$CDM$_{\rm H}$}
\def\HI{\hbox{H~$\rm \scriptstyle I\ $}}

\def\HeII{\hbox{He~$\rm \scriptstyle II\ $}}

\def\NHI{N_{\rm HI}}

\def\bmed{b_{\rm med}}
\def\cm2{\,{\rm cm$^{-2}$}\,}
\def\kms{\,{\rm km\,s$^{-1}$}\,}
\def\kmsmpc{\,{\rm km\,s$^{-1}$\,Mpc$^{-1}$}\,}

\def\kel{\,{\rm K\ }}

\def\ltsima{$\; \buildrel < \over \sim \;$}
\def\lsim{\lower.5ex\hbox{\ltsima}}
\def\gtsima{$\; \buildrel > \over \sim \;$}
\def\gsim{\lower.5ex\hbox{\gtsima}}
\def\etal{{ et~al.~}}
\def\dks{d_{\rm KS}}
\def\Pks{P_{\rm KS}}
\def\AutoVP{\small AutoVP\normalsize}
\def\SPECFIT{\small SPECFIT\normalsize}
\def\AutoVPbf{\small {\bf AutoVP}\normalsize}
\def\SPECFITbf{\small {\bf SPECFIT}\normalsize}
\def\AutoVPfc{\footnotesize AutoVP\small}
\def\SPECFITfc{\footnotesize SPECFIT\small}

\def\ana{A\&A}
\def\aj{AJ}
\def\apj{ApJ}
\def\apjs{ApJS}

\def\mn{MNRAS}
\def\mnras{MNRAS}

\journal{Preprint-01}

\input epsf

\begin{document}

\title{Hydrodynamical simulations of the Ly$\alpha$\ forest:\ data
comparisons }

\author[A. Meiksin et al.]{Avery Meiksin${}^{1}$, Greg Bryan${}^{2, 3}$,
Marie Machacek${}^{2, 4}$ \\
${}^1$Institute for Astronomy, University of Edinburgh,
Blackford Hill, Edinburgh\ EH9\ 3HJ, UK \\
${}^2$Department of Physics, MIT,
77 Massachusetts Avenue, Cambridge, MA 02139 , USA \\
${}^3$Hubble Fellow \\ 
${}^4$On leave from Department of Physics, Northeastern University,
Boston, MA, USA}

\pubyear{2001}

\maketitle

\begin{abstract}
Numerical hydrodynamical simulations are used
to predict the expected absorption properties
of the \Lya forest for a variety of Cold Dark Matter
dominated cosmological scenarios:\ CHDM, OCDM, \LCDMns, SCDM, and tCDM.
Synthetic spectra are constructed duplicating the resolution,
signal-to-noise ratio, and wavelength coverage of several published high
resolution spectra, and their statistical properties compared
on the basis of the flux distribution of the spectra,
the distribution of coefficients in a wavelet decomposition
of the spectra, and the distributions of absorption line profile parameters.
Agreement between the measured and predicted cumulative distributions is found
at the few to several percent level. The best-fitting models to the flux
distribution correspond to normalizations on the scale of the cosmological
Jeans length of $1.3<\sigma_{\rm J}<1.7$ at $z=3$. No single model provides a
statistically acceptable match to all the distributions. Significantly
larger median Doppler parameters are found in the measured spectra
than predicted by all but the lowest normalization models (CHDM and tCDM),
which provide poor fits to the flux distributions. The discrepancy in Doppler
parameters is particularly large for absorption systems optically thin at the
\Lya line-centre.
This may indicate a need to introduce additional energy injection throughout
the intergalactic medium, as may be provided by late \HeII reionization
($z_{\rm HeII}\simeq3.5$) or supernovae-driven winds from young galaxies,
and/or a larger baryon fraction than given by recent determinations of the
deuterium abundance within the context of standard Big Bang Nucleosynthesis.
The models require a hydrogen ionization rate at redshifts $1.7<z<3.8$ within
a factor of 2 of that predicted from QSOs alone as the sources of the UV
photoionization background, although with a slower rate of decline with
redshift at $z>3.5$ than predicted from current QSO counts. Principal
systematic uncertainties in comparing the models with the observations are the
setting of the continuum level of the QSO spectra and the prevalence of metal
absorption lines, particularly at $z<3$.
\end{abstract}


\section{Introduction}
\label{sec:introduction}

The successful recovery of the observed statistical properties of the
\Lya forest by numerical simulations in the context of various cold dark
matter (CDM) dominated cosmologies suggests that the gravitational instability
scenario provides a broadly accurate description of the development and
evolution of the structure of the intergalactic medium (IGM)
(Cen \etal 1994; Zhang, Anninos, \& Norman 1995; Hernquist \etal 1996;
Zhang \etal 1997; Bond \& Wadsley 1997; Theuns, Leonard \& Efstathiou 1998a).
The simulations show that nearly the entire IGM fragments
into filaments, sheets, and fluctuations in underdense minivoids, all
of which give rise to the absorption lines comprising the \Lya forest detected
in the spectra of Quasi-Stellar Objects (QSOs) (Cen \etal 1994;
Miralda-Escud\`e \etal 1996; Zhang \etal 1998). By analysing synthetic spectra
drawn from the simulations, it has been demonstrated that the measured flux
distributions, neutral hydrogen column density distributions, and the
evolution in the number of absorption lines per unit redshift are reasonably
well accounted for by the complex web of interconnecting structures found in
the simulations.

The structure of the \Lya forest is intermediate in complexity between
that of the Cosmic Microwave Background (CMB) and the large-scale
distribution of galaxies. It may be reasonably hoped that the physics
of gravitational instability and hydrodynamics alone, once a model for
the reionization of the IGM is assumed, is adequate for describing the
measurements of the \Lya forest. In this sense, the \Lya forest represents
a bridge between the relatively straightforward mechanisms believed
responsible for the formation of the CMB fluctuations (given an input
cosmological model and linear power spectrum), and the complex physical
processes involved in galaxy formation. For this reason, the \Lya forest
offers a unique forum for testing models of structure formation.

The Keck HIRES (Hu \etal 1995; Lu \etal 1996; Kirkman \& Tytler 1997) and,
more recently, the VLT UVES (Cristiani \& D'Odorico 2000) have provided high
resolution (6--9\kms), high signal-to-noise ratio (50--100 per pixel)
observations of the \Lya forest that resolve the structure of the absorption
features, and have dramatically improved the precision with which model
predictions may be compared with the measured statistical properties of the
\Lya forest. The high precision of the data has elicited similar precision in
the predictions from numerical simulations. Comparisons between the model
predictions and the observations have been effected by several groups by using
the simulation results to synthesize spectra that mimic the pixelization and
typical noise characteristics of the measured spectra. Adopting this approach,
Dav\'e \etal (1997) compare the predictions of their standard CDM (SCDM)
simulation for the distributions of \HI column densities and Doppler
parameters, as derived from Voigt profile absorption line fitting to
synthesized spectra, with the distributions reported by Hu \etal (1995);
Rauch \etal (1997) compare the predictions of an SCDM model and a flat CDM
model with a non-vanishing cosmological constant (\LCDM) for the
distribution of pixel fluxes with distributions measured from several Keck
HIRES QSO spectra; Theuns \etal (1998b, 1999) compare the predictions of an
open CDM (OCDM), SCDM, and \LCDM simulations for the distributions of
absorption line parameters with those reported by Hu \etal (1995), Lu \etal
(1996), and Kim \etal (1997); Weinberg \etal (1999) provide a preliminary
low-resolution comparison between the predicted flux per pixel distributions of
several cosmological models with that measured in a sample of Keck HIRES
spectra; and McDonald \etal (2000b) compare the prediction of the \LCDM
simulation of Miralda-Escud\'e \etal (1996) for the distribution of pixel
fluxes with the measured distribution averaged over several HIRES QSO spectra.
(In addition to mimicking the pixelization of the HIRES, Rauch \etal 1997,
Theuns \etal 1998b and McDonald \etal 2000b also incorporate the effect of the
spectral resolution of the HIRES in their synthetic spectra.) Because of the
required small box sizes of the simulations, the synthetic spectra are much
shorter than those measured, which cover a fairly broad range in redshift
(typically $\Delta z\approx0.5$). This limitation has necessitated several
approaches that degrade the precision of the comparison
between the simulation predictions and the data:\ assuming that an average
distribution measured over a range of redshifts is the same as the distribution
at the average redshift (Dav\'e \etal 1997; Theuns \etal 1998b, 1999;
McDonald \etal 2000; Machacek \etal 2000);
correcting the measured optical depths according to an assumed evolutionary
rule (Rauch \etal 1997, who assumed the pixel optical depths scale as
$\tau(z)\propto(1+z)^{4.5}$); or comparing
the predicted and measured distributions over broad redshift bins
(McDonald \etal 2000).

In addition to the tests above, there has been a parallel
effort to compare the predictions of the simulations for the flux
power spectrum (Croft \etal 2000 and references therein; McDonald
\etal 2000b) and auto- and cross-correlation functions (Zuo \& Bond
1994; Miralda-Escud\'e \etal 1996; McDonald \etal 2000b) with the
observations. These approaches target the measurement of
large-scale clustering (low $k$) properties of the IGM and,
implicitly, of the underlying dark matter, more than the smaller scale
(high $k$) absorption properties of the \Lya forest (although the two
are related), as a means for constraining models of large-scale structure
formation. In this paper we concentrate on the smaller
scale properties of the IGM and make no further mention of these
complementary statistics.

The comparisons between the simulations and data on the basis of the flux
and \HI column density distributions have consistently
produced very good agreement. An outstanding difficulty is the somewhat
poorer agreement found for the distribution of Doppler parameters. The
simulation convergence tests of Theuns \etal (1998b) and Bryan \etal (1999)
show that as the resolution of the simulations increases, the absorption
lines tend to narrow, although the trend need not be strictly monotonic
(Meiksin \& White 2001). As a consequence, the SCDM, OCDM and \LCDM simulations
predict smaller Doppler parameters than measured (at least when restricted to
$\NHI>10^{13}$\cm2 systems), without the introduction of additional heating
of the IGM (Theuns \etal 1999; Bryan \& Machacek 2000; Machacek \etal 2000;
Theuns, Schaye \& Haehnelt 2000). The improved agreement found by Machacek
\etal (2000) between the predictions of a tilted CDM (tCDM) model and the
distribution of Doppler parameters reported by Kim \etal (1997), however,
suggests that the case for disagreement may depend on the assumed cosmological
model.

While the concordance between the simulation predictions and measurements of
the \Lya forest has been good, statistical quantifications of the level of
agreement have generally been absent, rendering it difficult to assess the
importance of discrepancies, as in the Doppler parameter distributions. It also
leaves open the more general question:\ are the predictions of the numerical
simulations consistent with the observations at a statistically acceptable
level? It is a central goal of this paper to address this question. Progress
toward an answer has recently been made by McDonald \etal (2000b), who quantify
the agreement between the pixel flux distribution for a \LCDM simulation and
the mean measured distribution based on several HIRES spectra. To do so,
however, they split the flux values into several (21) bins and use the $\chi^2$
statistic as a basis for comparison, finding agreement between the simulated
and measured distributions for $z>3$, but not at $z=2.41$. Given the high
resolution and wide wavelength range of the spectra, a far greater number of
degrees-of-freedom are available, permitting more stringent tests.
For a typical pixel number of $N\approx10^4$, it should be possible to
measure the cumulative flux distribution to a precision of
$1/N^{1/2}\approx0.01$, or perhaps a few times this allowing for correlations
in the flux values of neighbouring pixels. No group has fully exploited the
information available from Voigt profile fitting, which typically produces
several hundred lines per spectrum.

A principal theme of this paper is to determine how closely the simulation
predictions match the statistics of the \Lya forest, as extracted directly
from the measured spectra, and to determine the limitations involved, both
observational and theoretical. To do so, we use a suite of simulations for a
variety of cosmological models to construct synthetic spectra designed to
match the resolution, pixelization, wavelength coverage, and noise properties
of a set of measured spectra published in the literature. Both the
synthesized and measured spectra are analysed using identical software to
minimize any differences due to the arbitrary nature of the algorithms. In this
way, we are able to determine how sensitive the statistics of the absorption
features are to the properties of the spectra, including data-reduction
systematics, like the uncertainty in the continuum-fitting. We employ
several statistical tools:\ the distribution of flux per pixel, the
distributions of \HI column density and Doppler parameters, and a novel
approach based on the Discrete Wavelet Transform (DWT) of the data
(Meiksin 2000).

This paper differs from previous work in several important regards:\ 1.\ We
simulate the full wavelength coverage of the measured spectra (between \Lya and
\Lyb), to the wavelength precision of the spectra, incorporating the
specific noise characteristics of each spectrum. 2.\ We consider a broader
range of statistical tests, and employ identical software to analyse both the
simulations and the measured spectra. 3.\ We introduce
statistical quantifications of the differences between the model predictions
and the observations, including analyses of the probability distributions of
the tests employed. 4.\ We consider a broader range of cosmological models than
has been done previously in the context of the full set of tests considered.
5.\ Lastly, our fluid computational method differs fundamentally from those
employed in most of the other simulations. While most use Smoothed Particle
Hydrodynamics (SPH) with artificial viscosity to solve for the hydrodynamics
part of the simulations, we use an Eulerian (in the co-moving frame)
finite-difference shock-capturing scheme employing the Piecewise Parabolic
Method (PPM). The exception is the \LCDM model of Miralda-Escud\'e \etal
(1996), who also used an Eulerian shock-capturing scheme, based on the total
variation diminishing scheme. We believe it important to consider simulations
using different numerical schemes, particularly when making high precision
comparisons with the data.

Our ultimate goal, toward which this paper is directed, is to
determine whether or not the data and the simulations are of adequate
accuracy to distinguish between differing cosmological models and
reionization scenarios. If so, the simulations should provide a
valuable independent tool for testing models of structure formation as
well as probing the physical state of the gas and the energetic
processes that accompanied the formation of early sources of
photoionization, whether QSOs, galaxies, or smaller stellar systems.

The paper is organized as follows. In the next section we describe the
methods used to compare the simulation results with the measured spectra. In
\S III, the analysis results for seven QSO spectra are presented. A discussion
of the results follows in \S IV. Our conclusions are summarised in \S V. 
Various tests of the analysis procedures and absorption line fitting
algorithms are presented in the two appendices.

\section{Analysis Methods}
\label{sec:analysis}

\subsection{Models and spectra}
\label{subsec:spectra}

The models used in the paper are described in detail in Machacek \etal
(2000) (M00) and summarised in Table~\ref{tab:model_par}. All the
models considered are in the context of CDM dominated cosmologies. The
following six models were simulated using Kronos (Bryan \etal 1995), a
combined Particle Mesh (PM) $+$ PPM gravity-hydrodynamics code:\ a standard
critical-density flat CDM model (SCDM), two flat CDM models with a
non-vanishing cosmological constant and low (\LCDMLns) and high
(\LCDMHns) baryon densities, an open CDM model (OCDM), a tilted flat
CDM model (tCDM), and a flat critical-density mixed dark matter model
with a hot component added to the CDM (CHDM). All the power spectra
have been normalized to be consistent with the abundances of rich
clusters (White, Efstathiou \& Frenk 1993; Bond \& Myers 1996; Eke, Coles
\& Frenk 1996; Pierpaoli, Scott \& White 2000), as the scales involved are near
the scales relevant to the formation of the structures giving rise to the \Lya
forest. In addition, the tilt adopted in the tCDM model was chosen also to
match the normalization on large scales (although only marginally) as
determined from the COBE measurements of the CMB (Bunn \& White 1997). The
\LCDMH model was designed to yield larger Doppler parameters than the \LCDML
model in an attempt to better match the measured values. Accordingly,
in addition to increasing the baryon density,
we have  boosted the rate of \HeII heating by QSO sources by a
factor of 1.8 to mimic the sudden onset of \HeII photoionization at
late times (Bryan \& Machacek 2000). There is some uncertainty as to the best
value of the baryon density $\Omega_bh^2$ to adopt, as there is some spread
depending on the method by which it is measured. Measurements based on
deuterium abundances within the context of standard Big Bang Nucleosynthesis
(BBN) (Burles \& Tytler 1998; O'Meara \etal 2000) yield lower values than the 
best estimates based on measurements of the acoustic peaks in the CMB
(Netterfield \etal 2001; Pryke \etal 2001; Stompor \etal 2001).
The values adopted for all the models are consistent with
the limits of Copi, Schramm \& Turner (1995) based on a variety of
observational constraints. The value used for the \LCDMLns,
OCDM and SCDM models is at the 3$\sigma$ lower limit of Burles \& Tytler (1998)
and O'Meara \etal (2000) based on measurements of the deuterium abundances in
intergalactic gas clouds, and assuming no destruction of primordial deuterium.
(Absorption systems with low metallicity were specifically chosen by these
groups to ensure the effects of stellar processing were minimal, but a small
amount of processing would in principle permit a lower value for
$\Omega_bh^2$.) The value for the \LCDMH model agrees with the best estimate of
O'Meara \etal (2000) and Pryke \etal (2001). The value used for the CHDM and
tCDM models is midway between the best estimates of O'Meara \etal (2000) and
Stompor \etal (2001).

The initial data for all models except the CHDM and \LCDMH models were
generated using COSMICS (Bertschinger 1995) with the BBKS transfer
function (Bardeen \etal 1986) to compute the starting redshifts and
the unconstrained initial particle positions and velocity
perturbations appropriate for each model. For the CHDM model we used
CMBFAST (Seljak \& Zaldarriaga 1996) to generate the initial
conditions. For the \LCDMH model the power spectrum was based on the
analytic fits of Eisenstein \& Hu (1999).

We use the synthetic spectra described in M00. The spectra are resampled
according to the pixelizations given in Table~\ref{tab:qsos}, and the spectral
resolution mimicked by smoothing the spectra with a gaussian of the indicated
widths. Noise was added according to the varying signal-to-noise ratio across
each measured spectrum, as described below.

In order to construct spectra with the identical wavelength coverage as the
observed spectra, the simulation results from a broad range in redshifts must
be used. Since the simulation results are saved at only integral redshift
values between $z=1$ and $z=5$ (except for the \LCDMH model, for which the
data are saved more frequently), it is necessary to interpolate between
redshifts. (We do not use the $z=1$ results since the fundamental mode across
the simulation volume has grown too large by this time to be certain of the
reliability of the results.) To assist in the interpolation, we take advantage
of the redshift scaling relations of the gas properties found by M00. Because
the gas is in photoionisation equilibrium, the optical depth at
any given wavelength scales like $\tau(\lambda)\propto n^2\alpha_A(T)/
[\Gamma(z)H(z)]$, where $n$ is the gas density, $\alpha_A(T)$ is the (Case A)
radiative recombination coefficient at gas temperature $T$, $H(z)$ is the
Hubble constant at redshift $z$, and $\Gamma(z)$ is the incident rate of
ionising photons per neutral atom. Since the morphologies and overdensities of
the overdense structures evolve only slowly in the comoving frame and $T$
scales like the square-root of the gas overdensity (Zhang \etal 1998), and
since $\alpha_A(T)\propto T^{-0.7}$ (approximately), we may expect the
combination $n^2\alpha_A(T)$ to scale like a power of the mean gas density
$\bar n(z)$, which itself scales as $(1+z)^3$. Accordingly, we scale the
optical depths according to $\tau(z)\propto(1+z)^\alpha/[\Gamma(z)H(z)]$. The
value $\alpha=6$ is expected for a homogeneously expanding medium of fixed
temperature. We find that in general $\alpha$ is model-dependent.
For any particular cosmological model, we determine $\alpha$
between two redshifts $z_1$ and $z_2$ by finding the value which
best predicts the mean optical depth, defined by
$\bar\tau_\alpha\equiv -\log\langle\exp(-\tau)\rangle$ averaged over the
spectra, at $z=z_2$ using the spectra at $z=z_1$. (In principle, a different
value for $\alpha$ may be found by reversing the roles of $z_1$ and $z_2$. We
checked this and found in all cases the same value for $\alpha$ was derived.)
The values found are provided in Tables~\ref{tab:BT97_rescale} and
\ref{tab:KT97_rescale}. We allow for an overall rescaling of the radiation
field by multiplying the optical depths by a scale factor $s$. (The factor $s$
could also be interpreted as a rescaling of the baryon density, although this
would entail a change in the gas temperature, which we do not allow for.)
The UV background radiation field adopted in the simulations evolves according
to the model of Haardt \& Madau (1996) predicted for QSOs as the sources of the
radiation, and assuming an intrinsic QSO spectral index $\alpha_Q=1.5$. The
radiation field is turned on at $6<z_{\rm ion}<7$. We additionally accommodate
an alternative evolution in the radiation field by allowing for a further
evolution factor $(1+z)^p$ multiplying the Haardt \& Madau rate
$\Gamma_{\rm HM}(z)$. Thus we predict $\tau$ at any
given wavelength $\lambda$ at redshift $z$ based on the value at redshift
$z_1$ at wavelength $\lambda(1+z)/(1+z_1)$ according to
\begin{equation}
\tau(z)=s\tau(z_1)\left(\frac{1+z}{1+z_1}\right)^\alpha
\left(\frac{1+z_{\rm ref}}{1+z}\right)^p
\frac{\Gamma_{\rm HM}(z_1)H(z_1)}{\Gamma_{\rm HM}(z)H(z)},
\label{eq:rescale}
\end{equation}
where $z_{\rm ref}$ is a reference redshift introduced for $p\ne0$. The values
for $s$, $p$, and $z_{\rm ref}$ adopted for different cosmologies and required
to match the measured spectra are provided below.

The interpolation to intermediate redshifts to cover the wavelength range of a
particular measured spectrum is performed by generating spectral segments
from the simulations, each a fixed redshift interval $\Delta z=0.1$ in length,
scaling the optical depth in each segment according to
equation~\ref{eq:rescale}, and then effectively piecing the segments together
to match the wavelength range
of the measured spectrum. In order to interpolate between two data dumps
at redshifts $z_1$ and $z_2$, we compute the distribution functions
$f(x;z,z_d)$ for any given statistic $x$ within a redshift interval
$(z, z+\Delta z)$ by first extrapolating the optical depths from a data dump
at $z_d$ to $z$, and then computing the final distribution function for a
redshift interval $z_b<z<z_a$ according to
\begin{equation}
f(x) = \sum_i\Delta z \left[w_{2,i}f(x;z_i,z_1)+w_{1,i}f(x;z_i,z_2)\right],
\label{eq:interp}
\end{equation}
using the linear weights $w_{j,i}=\vert(z_i-z_j)/(z_2-z_1)\vert$, and
incrementing $z_i$ from $z_b$ to $z_a$. We find this
procedure recovers the distribution functions at intermediate redshifts to
very high accuracy, as is shown in Appendix~\ref{app:tests}.

In all cases, the wavelength coverage $z_b<z<z_a$ corresponds to a
range between \Lya and \Lyb in the QSO restframe to avoid confusing
\Lyb absorption with \Lya absorption. In practice the ranges are usually
shortened to a smaller interval, depending on the wavelength coverage of
the spectrum and to avoid the local photoionizing influence of the QSO (the
proximity effect).

A major uncertainty in comparing with the measured spectra is the setting of
the continuum level. Normally peak flux points are used to determine
the level of the
continuum in the \Lya forest region of the spectrum. In regions where the flux
never recovers to the true continuum, however, this procedure will result in an
underestimate of the true continuum level. Without an automated procedure for
inferring the continuum within the forest, it is difficult to gauge the effect
of the uncertainty in the continuum level. We find that making no correction to
the continuum frequently results in a mismatch in the flux distribution for
flux values near unity (in units of the continuum). We model errors in the
continuum level by allowing for a shift in its overall value according to:
\begin{equation}
F_i = \frac{F_i^{(0)}}{1-\omega},
\label{eq:offset}
\end{equation}
where $F_i^{(0)}$ is the flux predicted by the numerical model at pixel $i$,
and $F_i$ is a corrected flux allowing for a downward shift in the continuum
level by the fractional amount $\omega$, applied uniformly across the spectrum.
This is a crude approximation to the actual errors likely to occur. We adopt it
for its simplicity in order to assess the impact of possible continuum errors.
For a more accurate comparison, it would be necessary to re-analyse the raw
data and offset the simulation spectra at each pixel following a procedure
consistent with that of the observers.

The fitting of absorption lines to the spectra requires a noise value for
each pixel. We generate a noise matrix from the $1\sigma$ noise levels in
the observed spectra, averaging the noise in several flux bins over redshift
intervals of width $\Delta z=0.1$. We then draw a noise value assuming the
noise to be gaussian distributed with a standard deviation corresponding to
the matrix element of the redshift and flux of the pixel. This allows us to
model a varying signal--to--noise ratio across a spectrum, and allows for a
correlation between flux and noise. We have verified that the statistical
properties of the spectra are not very sensitive to the noise model:\ averaging
over redshift bins but not allowing for a correlation between flux and noise
level yields very similar results.

\subsection{Statistical analysis tools}

There are several statistics that may be used as a basis of comparison between
the models of the \Lya forest and the observed spectra. The most fundamental is
simply the distribution of flux per pixel. A high level of degree of accuracy
in the cumulative flux distribution may be achieved because of the large number
of pixels in the spectra. The measured and predicted cumulative distributions
may be compared using the Kolmogorov--Smirnov (KS) test, for which the expected
value of the difference (in absolute value) $\dks$ between the
distributions for $N$ measurements is $\bar\dks\simeq0.9/N^{1/2}$, and $\dks$
should exceed $1.6/N^{1/2}$ only 1 per cent of the time. For a typical
number of pixels $N\sim10^4$, the agreement between the predicted and
measured cumulative flux distributions should then be $\dks\approx0.01$.
The probability distribution is based on the assumption that the measurements
are statistically independent. Because the absorption features will introduce
correlations between neighbouring pixel fluxes (as will the resolution of the
spectrographs), this assumption is formally violated. As is shown in
Appendix~\ref{app:tests}, the correlations effectively tend to reduce the
number of independent pixels by a factor of 3--5. Even allowing for this,
the test still provides a powerful means of quantifying the statistical
agreement between the model predictions and the measured spectra.

Additional constraints are provided by the fluctuations in the
spectra, which are related both to the underlying baryon density
fluctuations and to the thermal and velocity widths of the associated
absorption lines. We quantify the fluctuations in two ways. The scales
of fluctuation of the spectral features may be directly characterised
using the Discrete Wavelet Transform, which quantifies the changes in
a given spectrum on being smoothed from one velocity resolution scale to
another (Meiksin 2000). A multiscale analysis based on the Daubechies
wavelets effectively performs a smoothing of the spectrum over
successive doublings of the pixel width, weighted such that the
spectrum is decomposed into orthogonal components corresponding to
different velocity scales. (We use the Daubechies wavelet of order 20.)
A wavelet analysis offers several advantages over the more traditional
Voigt-profile absorption line analysis:\ 1.\ the method is objective, in that
it does not require making arbitrary decisions regarding the deblending of
absorption lines or the acceptability of a fit; 2.\ the method produces a
large number of wavelet coefficients, $N/2^L$ at the resolution level
$L$ for $N$ total pixels, corresponding to a typical
$\dks\approx0.01\times2^{L/2}$, thus enabling stringent statistical constraints
to be set; 3.\ the coefficient distributions are insensitive to noise
(Meiksin 2000); 4.\ the analysis is extremely fast. Monte Carlo
simulations show that the wavelet coefficients are only weakly correlated, so
that they may be treated as statistically independent (Meiksin 2000). There
are, however, some disadvantages as well:\ 1.\ the light fluctuations are
quantified only over successive doublings of the pixel size, allowing only
coarse velocity resolution in the statistical quantification of the
fluctuations; 2.\ the method is indifferent to the origin of the
fluctuations, and in particular does not distinguish directly between the
effects of \HI column density and Doppler broadening, which may especially
become mixed in the presence of substantial line-blending. For these reasons,
we perform Voigt profile analyses as well.

We utilise \AutoVP\ (Dav\'e \etal 1997) to decompose the spectra into a set
of Voigt absorption line profiles. Typically 300--500 lines are found per
spectrum, providing a precision of $\dks\approx0.05$ in the cumulative
distributions of $\NHI$ and $b$. In order to determine the sensitivity of the
resulting distributions of absorption line parameters to the line analysis
method, we have also developed a separate algorithm called
\SPECFIT. Rather than searching for absorption maxima, and adjusting
the number of lines required for a good fit, as does \AutoVP,
\SPECFIT\ first splits the spectra into segments bordered by flux
values above that corresponding to a minimum optical depth threshold
$\tau_{\rm min}$, and then searches for inflection points in
noise-filtered representations of the data in each segment in order to
identify the locations of candidate absorption lines. The candidate
lines are then fit to the original spectrum. While less sophisticated
than \AutoVP, \SPECFIT\ operates a factor of 50-100 faster with a
comparable level of success in the fits, except that it sometimes
misses the weakest optically thin features ($\tau<\tau_{\rm
min}$). The two methods are compared in Appendix~\ref{app:lpcomp}.

In the following, we shall generally regard the $3\sigma$ level (a KS test
probability of $\Pks=0.001$) as the threshold for rejecting a predicted
distribution compared with the measured using the KS test.

\subsection{Data}

We compare the simulation results with six Keck HIRES spectra and one VLT UVES
spectrum that have been published in the literature. The quasars, the redshift
range used in the comparison, the pixel resolutions of the spectra, and the
spectral resolutions, are summarised in Table~\ref{tab:qsos}. All the spectra
were normalized to a unit continuum level by the observers.


\section{Results}
\label{sec:results}

In this section, we present comparisons between the simulations and the data
as quantified by the flux per pixel distributions and the distributions of
wavelet coefficients and absorption line parameters. Because of differences in
the characteristics of the spectra and the dispersion in the statistics
extracted from the measured spectra, we present the results for each spectrum
separately. The reader interested primarily in the implications of the results
may choose to skip this section and proceed to Section~\ref{sec:discussion}.

The amplitudes of the distributions found in the models cannot be used as a
basis of comparison since the optical depths may be arbitrarily rescaled for
any individual model by the ionization bias factor $b_{\rm ion}\propto
\Omega_b^{1.6}/\Gamma$ (Hui \& Gnedin 1997; Croft \etal 1997).
(We note the temperature--density relation found by Zhang \etal 1998 for
systems in the column density range $12.5<\log_{10}\NHI<14.5$ is consistent
with this dependence.) The dependence on $\Omega_b$ is based on an approximate
polytropic relation between density and temperature found to hold at low
density (Meiksin 1994; Hui \& Gnedin 1997), corresponding to the low and
moderate column density systems which dominate the absorption in the spectra.
The relationship, however, is not valid for higher column density systems
($\log_{10}\NHI>15$) because these arise predominantly in high density regions
(Zhang \etal 1998), for which the relationship inverts (Meiksin 1994;
Theuns \etal 1998b). The scaling
also neglects the effects of changes in the gas temperature on absorption
profiles that will result from a change in the baryon density, or of any
associated pressure changes. For these reasons, it is most conservative to view
a change in $b_{\rm ion}$ as a change in $\Gamma$, which has little effect on
the gas temperature or pressure. It is important to
normalize all the models consistently with any given measured QSO
spectrum before comparing the shapes of any of the distributions with
those measured. This may be accomplished in a variety of ways. We do so by
matching the mean \HI optical depth $\bar\tau_\alpha$ in each
simulation to the measured intergalactic \HI optical depth over the
comparison redshift interval. Since the resulting model flux
distributions will often mismatch the observed distributions for flux
values near unity, we will also allow for a constant offset in the
continuum as given by equation~\ref{eq:offset}, while keeping the value of
$\bar\tau_\alpha$ fixed at the observed value. In some instances it is
also found necessary to introduce additional evolution in the UV
radiation background to that adopted in the simulations.

\subsection{Q1937$-$1009}

\subsubsection{Flux distribution}

\begin{figure}
\begin{center}
\leavevmode \epsfxsize=3.3in \epsfbox{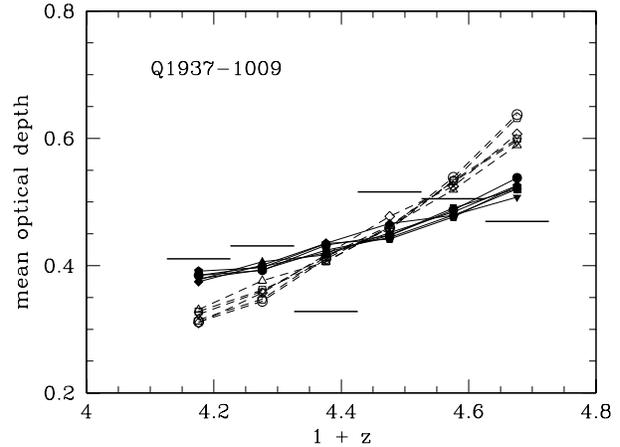}
\end{center}
\caption{Redshift dependence of mean optical depth $\bar\tau_\alpha$ for
Q1937$-$1009 (horizontal bars). The open symbols show the values for
$\bar\tau_\alpha$ found in the simulations based on the evolution of the UV
radiation background as determined from QSO sources by Haardt \& Madau (1996).
The filled symbols show the values for $\bar\tau_\alpha$ after allowing for an
alternative redshift dependence of the UV radiation background. The symbols
are:\ CHDM (circles), \LCDML (squares), \LCDMH (diamonds), OCDM (triangles),
SCDM (inverted triangles), tCDM (pentagons).}
\label{fig:BT97_taua}
\end{figure}

We investigate Q1937$-$1009 (Burles \& Tytler 1997) most thoroughly in order
to assess the magnitude of some of the possible systematics in our analyses.
Over the redshift range $3.126 < z < 3.726$, the mean optical depth in the
measured spectrum is $\bar\tau_\alpha=0.44$. The mean optical depths in
redshift intervals $\Delta z=0.1$ wide are shown in Fig.~\ref{fig:BT97_taua},
along with estimates from each of the models. The optical depth per
pixel has been rescaled according to equation~\ref{eq:rescale} for each model,
using the parameter values given in Table~\ref{tab:BT97_rescale}. We note that
the redshift dependences for the optical depth per pixel found in the
simulations are generally steeper than the dependence expected in a
homogeneously expanding universe: $\tau(z)\propto(1+z)^6/H(z)$ (assuming
constant $\Gamma$). By contrast, the measured values of the mean optical depth
$\bar\tau_\alpha$ are nearly flat with redshift. Formally they are well fit by
$\bar\tau_\alpha=a(1+z)^b$, with $a=0.022_{-0.015}^{+0.044}$ and
$b=2.0\pm0.74$, where the standard deviation of the measured values of
$\bar\tau_\alpha$ is adopted as a measurement error for each value.
The models predict a near doubling of the mean optical depth over
the same redshift interval, corresponding to $b=6$, significantly steeper than
measured. It is evident that an alternative evolution rate to that of the
simulations is required to match the redshift dependence of the mean optical
depths. To match the observed evolution, we adopt the additional rate exponents
$p$ for $z_{\rm ref}=3$ (cf equation~\ref{eq:rescale}) as given in
Table~\ref{tab:BT97_rescale}. The resulting distributions of $\bar\tau_\alpha$
for the various models are shown in Fig.~\ref{fig:BT97_taua}.

\begin{figure}
\begin{center}
\leavevmode \epsfxsize=3.3in \epsfbox{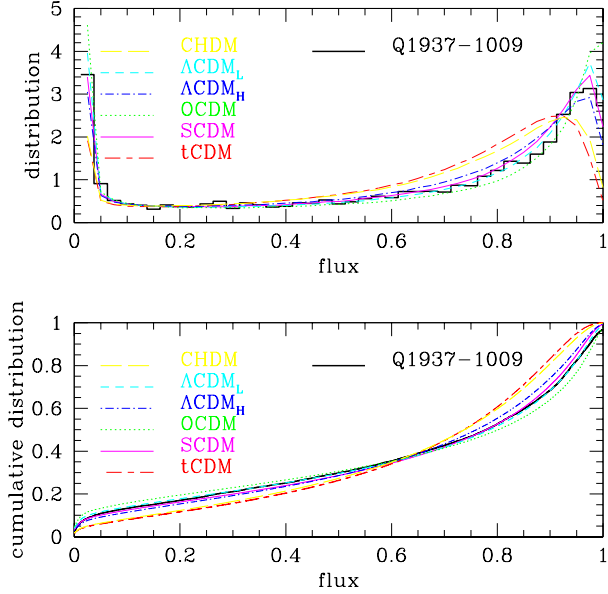}
\end{center}
\caption{Comparison between the measured flux distribution of Q1937$-$1009 and
the model predictions. Particularly good agreement is found for the \LCDML
model, although it is formally rejected by the KS test. (The measured
distribution has been rebinned in the upper panel for clarity.)}
\label{fig:BT97_fluxdist}
\end{figure}

The distributions of flux per pixel for the various models are compared with
the observed distribution in Fig.~\ref{fig:BT97_fluxdist}. (Flux values
exceeding unity result from noise fluctuations.) The agreement is generally
good, particularly for the \LCDML model. The models, however, all show some
disagreement at flux values near unity, and for this reason they are all
formally rejected by the KS test. For the \LCDML model, the
maximum difference (in absolute value) between the predicted and measured
cumulative distributions is $\dks=0.022$, corresponding to a formal acceptance
probability of $\Pks=10^{-4}$. The next best model is SCDM, which gives
$\dks=0.039$ and $\Pks=8\times10^{-14}$.

\begin{figure}
\begin{center}
\leavevmode \epsfxsize=3.3in \epsfbox{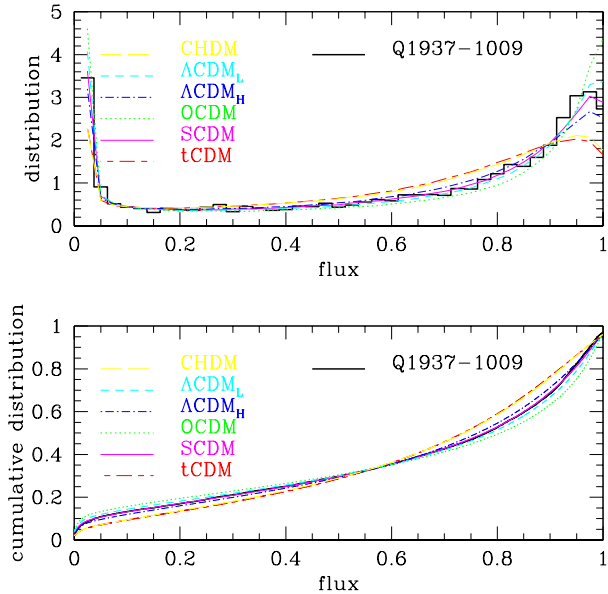}
\end{center}
\caption{Comparison between the measured flux distribution of Q1937$-$1009 and
the model predictions. A continuum offset is applied to each model to enforce
matching the cumulative distributions at flux values near unity. The
distributions tighten compared with Figure~\ref{fig:BT97_fluxdist}. The SCDM
model is now preferred over \LCDML. (The measured distribution has been
rebinned in the upper panel for clarity.)}
\label{fig:BT97_fluxdist_off}
\end{figure}

Allowing for the possibility that the continuum level in the observed spectrum
was set slightly low, resulting in overestimates of the normalized flux, we
increase the flux per pixel according to equation~\ref{eq:offset}, choosing
values for $s$ and $\omega$ that simultaneously preserve $\bar\tau_\alpha=0.44$
and match the measured flux distribution at flux values near unity. The
resulting flux distributions are shown in Fig.~\ref{fig:BT97_fluxdist_off},
using the values for $s$ and $\omega$ in Table~\ref{tab:BT97_rescale}. The
agreement in the cumulative distributions is now tightened, with the SCDM
model agreeing best with the observed distribution. For the SCDM
model, $\dks=0.016$, corresponding to a formal acceptance probability by the
KS test of $\Pks=0.011$. By comparison, the next best models are \LCDMH and
\LCDMLns, with $\dks=0.031$ and 0.033, respectively, both of which are strongly
rejected by the KS test, with respective acceptance probabilities of
$\Pks=4\times10^{-9}$ and $5\times10^{-10}$.

We recall that the KS test is based on the probability distribution of
$N^{1/2}\dks$, where $N$ is the number of independent measurements in
the sample. In Appendix~\ref{app:tests}, it is shown that the probability
distribution for this quantity based on the simulated spectra is broader
than the theoretical distribution, as may be expected if the flux values in
neighbouring pixels are correlated. Because of the limited number of
lines-of-sight drawn from the simulation volume, it is not possible to attach
meaningful probabilities to large values of $\dks$ directly from the
simulations. It is shown in Appendix~\ref{app:tests}, however, that the
probability distribution is close to that of the KS test for an effective
number of pixels reduced by a factor of 3--5 from the actual number. Allowing
for this reduction, the KS probabilities for the SCDM, \LCDMH, and \LCDML
models (with continuum offsets applied)
become, respectively, $\Pks=0.36-0.68$, 0.002--0.037, and 0.001--0.024,
still favouring SCDM, but not excluding the \LCDM models. The predictions of
the CHDM, OCDM and tCDM models, however, are still strongly rejected. We note
that not allowing for the continuum offset correction for \LCDML gives the KS
probability, allowing for the reduced effective number of pixels,
$\Pks=0.081-0.29$, so that the \LCDML model is perhaps no less viable than
the SCDM model. The difference in the probability value with and without a
continuum offset illustrates how susceptible $\Pks$ is to even a
small degree of uncertainty in the continuum level. We note
that while each of the \LCDML models disagrees with the measured flux
distribution more than does the SCDM model, the predicted flux distributions
bracket the measured distribution, so that an intermediate \LCDM model
may be expected to provide an even better match.

\subsubsection{Wavelet coefficient distributions}

\begin{figure}
\begin{center}
\leavevmode \epsfxsize=3.3in \epsfbox{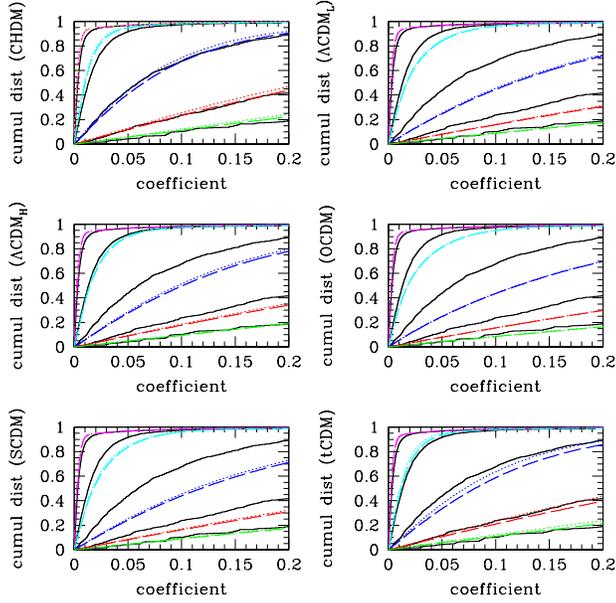}
\end{center}
\caption{Comparison between the measured wavelet coefficient cumulative
distributions of Q1937$-$1009 (solid curves) and the model predictions with
(dashed curves) and without (dotted curves) continuum offset corrections. The
curves from left to right correspond to the approximate velocity scales
4--8, 8--17, 17--34, 34--68, and 68--136\kms. The predictions of the CHDM model
agree best with the measured distributions.}
\label{fig:BT97_wc_off}
\end{figure}

The distribution of wavelet coefficients is shown in
Fig.~\ref{fig:BT97_wc_off}, both with no continuum offset applied and with
the offsets in Table~\ref{tab:BT97_rescale}. The models generally show poor
agreement with the measured distribution at the smallest velocity scale,
$\sim4-8$\kms. It is possible that this is due to an inadequancy in the
modelling of the spectral characteristics of the Keck HIRES in the synthetic
spectra. Better agreement with Keck data, however, was achieved by Meiksin
(2000) using similar modelling for a set of Monte Carlo realisations. It may be
that the disagreement results from limitations of the simulations. We discuss
this further in Section~\ref{sec:discussion} below. A discrepancy is also
found on the next scale of 8--17\kms, with half the models (\LCDML, OCDM
and SCDM) exhibiting
too great an amount of velocity structure (as indicated by the larger values of
the coefficients). The discrepancy between these models and the data suggests a
greater number of absorption features predicted by the simulations on these
scales than in the data. The lower normalization models, by contrast, predict a
smaller amount of structure than in the data, while the \LCDMH model is
intermediate and agrees best with the data (though it is formally rejected by
the KS test, with $\dks=0.05$ and $\Pks=6\times10^{-7}$). The disagreement at
the next velocity scale of 17--34\kms is generally much worse. These velocities
correspond to the typical Doppler parameters of the absorption features
(see below). Only the CHDM model provides an acceptable match to the data on
these scales; the other models predict far too much structure. The effect of
the continuum offset correction is generally negligible on all velocity scales,
except for CHDM and tCDM (which have the largest corrections), although it is
small even for these. For CHDM, the KS test gives on the scale 17--34\kms,
$\dks=0.041$ and $\Pks=0.030$ without the offset correction, and $\dks=0.035$
and $\Pks=0.083$ with the correction. The agreement between the models and the
data tends to improve for all the models at larger velocity scales.

\subsubsection{Absorption line parameter distributions}

\begin{figure}
\begin{center}
\leavevmode \epsfxsize=3.3in \epsfbox{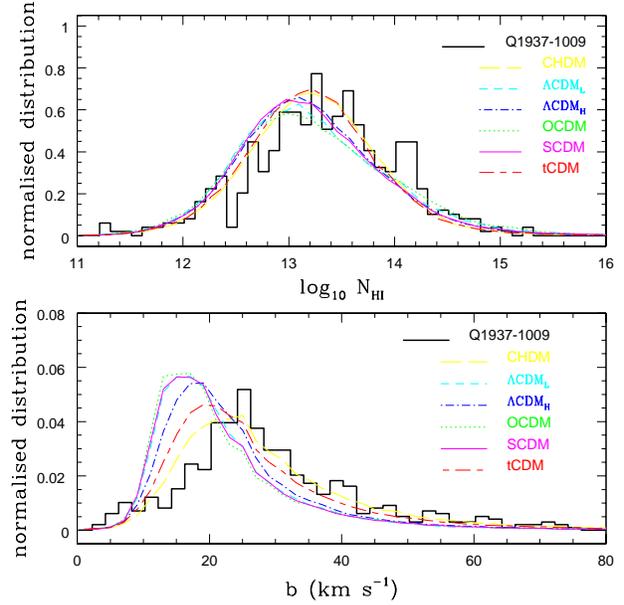}
\end{center}
\caption{Comparison between the measured absorption line parameter
distributions of Q1937$-$1009 and the model predictions using \AutoVP,
allowing for an offset in the continuum level. While good
agreement is generally found for the $\NHI$ distributions, the predicted
$b$ distributions are generally skewed toward lower $b$-values than measured.
(The measured distributions have been rebinned for clarity of presentation.)}
\label{fig:BT97_lpd_off}
\end{figure}

The \HI column density and Doppler parameter distributions resulting from
fitting absorption lines to the spectra using \AutoVP\ are shown in
Fig.~\ref{fig:BT97_lpd_off}. The continuum offsets of
Table~\ref{tab:BT97_rescale} have been applied. For each model, effectively 50
separate spectra were constructed from the simulation box corresponding to the
identical spectral range analysed in Q1937$-$1009. The profile analysis is
computationally expensive. For the range of column densities of
interest, it is adequate to fit the spectra using Doppler profiles rather than
full Voigt profiles. We find there is little difference in the resulting
distributions, while a factor of 2 is saved in the computing time.
Typically $2-3\times10^4$ features in total are found for each
model, requiring an analysis time of $\sim1-2\times10^5$ cpu seconds on a
Compaq XP900 with a 1GHz processor.

The agreement in the distribution of $\NHI$ between the data and all the models
is reasonably good, with the prediction of CHDM providing the best match, as
is found in the wavelet analysis. The predicted $b$ distributions for the
remaining models peak at substantially lower values than does the measured
distribution. For both the $\NHI$ and $b$ distributions, we find that the
probability distribution for $N^{1/2}\dks$, where $N$ is the number of
absorption lines, is narrower than the theoretical distribution, so that the
rejection probabilities provided by the KS test below are perhaps conservative.
We discuss this point further in Appendix~\ref{app:tests}.

\begin{figure}
\begin{center}
\leavevmode \epsfxsize=3.3in \epsfbox{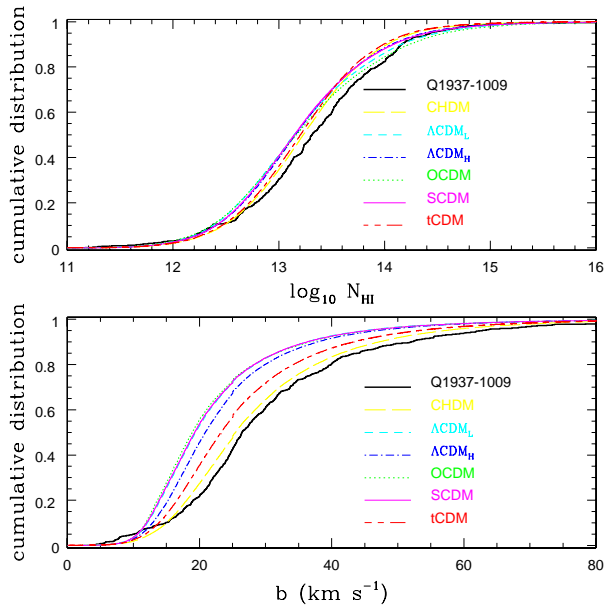}
\end{center}
\caption{Comparison between the measured absorption line parameter
cumulative distributions of Q1937$-$1009 and the model predictions using
\AutoVPfc, allowing for an offset in the continuum level. The predicted $\NHI$
distributions tighten about the measured distribution.
The CHDM model agrees best with the measured $\NHI$ and $b$ distributions.}
\label{fig:BT97_lpcd_off_avp}
\end{figure}

The cumulative distributions are shown in Fig.~\ref{fig:BT97_lpcd_off_avp}.
The measured $\NHI$ distribution is best matched by the CHDM model:\ the
maximum difference in the measured and predicted cumulative distributions is
$\dks=0.083$, with the associated formal KS acceptance probability
$\Pks=0.002$. The next best model is tCDM, with $\dks=0.10$ and
$\Pks=4\times10^{-5}$. The CHDM model also provides the best match to the
measured $b$ distribution, with $\dks=0.071$ and $\Pks=0.013$. (The remaining
models are very strongly rejected; the next best is tCDM with $\dks=0.16$
and $\Pks=4\times10^{-12}$.) It is clear that all the models other than CHDM
predict far too narrow lines. The \LCDMH model, which yields the largest
$b$-values of those models for which the predicted pixel flux distribution is
consistent with the measured, predicts a median Doppler parameter of
$\bmed=20.9$\kms. The measured median $b$ is
$26.4\pm0.9$\kms\footnote{The error in the median is $(0.5\pi/N)^{1/2}\sigma$,
where $\sigma$ is the standard deviation of the distribution of $b$-values, $N$
is the number of lines, and assuming the distribution of $b$-values is a
guassian. The data and the simulations show a long tail toward high values, and
so the error estimate should be modified accordingly (Kendall \& Stuart 1969).
We find that a lognormal distribution provides an acceptable fit to the
distribution of $b$-values derived by \AutoVPfc. Using the lognormal
distribution instead of a gaussian changes the error on the median by less
than one per cent.}, too large by $6\sigma$, but consistent with
$\bmed=25.5$\kms predicted by the CHDM model.

The narrowness of the absorption lines is reflected by the larger number of
systems found in the \AutoVP\ analyses of the simulation results:\ more lines
are required to recover the mean measured optical depth of the spectrum.
The \AutoVP\ analysis of the spectrum of Q1937$-$1009 yields a total of 495
absorption features. The expected numbers of lines predicted by the models are
for CHDM:\ 543; \LCDMLns:\ 592; \LCDMHns:\ 597; OCDM:\ 571;
SCDM:\ 618, and tCDM:\ 585, all of which except CHDM exceed the measured
number by at least $3\sigma$, assuming Poisson fluctuations.

The effect of the continuum offset is small in comparison with the variances
between the predicted and
measured $\NHI$ and $b$ distributions. For instance, for the \LCDML
model, which provides the best fit to the pixel flux distribution without a
continuum offset correction, the median Doppler parameter is 20.2\kms with no
offset correction and 19.4\kms with the correction. Both values are
inconsistent with the measured median $b$-value by over $6\sigma$. The KS test
formally strongly rejects the $\NHI$ and $b$ distributions, compared with the
measured distributions, for both cases.

\begin{figure}
\begin{center}
\leavevmode \epsfxsize=3.3in \epsfbox{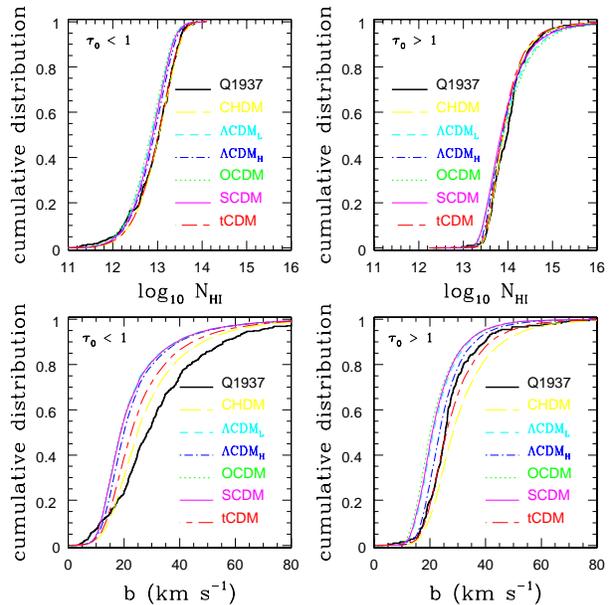}
\end{center}
\caption{Comparison between the measured absorption line parameter
cumulative distributions of Q1937$-$1009 and the model predictions using
\AutoVPfc, for subsamples of lines optically thin ($\tau_0<1$) and optically
thick ($\tau_0>1$) at \Lya line-centre. The agreement between the model
predictions and the data is best for the optically thick systems. All the
models predict much too low median $b$-values for the optically thin systems.}
\label{fig:BT97_lpcd_split_avp}
\end{figure}

Absorption features with a line-centre optical depth $\tau_0<1$ correspond to
structures that are underdense at $z=3$, while higher optical depth systems are
associated with overdense gas (Zhang \etal 1998).
Underdense gas at these redshifts is too rarefied to maintain thermal
equilibrium between photoionization heating and radiative losses, and so will
retain a partial memory of its thermal history (Meiksin 1994). Overdense gas is
more readily able to maintain thermal equilibrium. Because of this
physical difference, we split
the line samples into two subsamples at $\tau_0=1$ to determine if the
disagreements between the measured and predicted $\NHI$ and $b$ distributions
may be due to an incorrect thermal history rather than an incorrect
cosmological model. The effect of this split is shown in
Fig.~\ref{fig:BT97_lpcd_split_avp}. All of the models produce $\NHI$
distributions for $\tau_0>1$ consistent with the data. The best case is CHDM,
with $\dks=0.10$ and $\Pks=0.041$. The worst cases are the
SCDM and tCDM models, both with $\dks=0.14$ and $\Pks=0.002$, while the \LCDML
and \LCDMH models give, respectively, $\dks=0.12$, $\Pks=0.014$ and
$\dks=0.11$, $\Pks=0.018$. Only the tCDM model is able to
reproduce the $b$ distribution for $\tau_0>1$ ($\dks=0.06$, $\Pks=0.50$).
All the models predict a number of
absorption systems consistent with the measured number of 179 (CHDM:\ 149;
\LCDMLns:\ 205; \LCDMHns:\ 184; OCDM:\ 214; SCDM:\ 206; tCDM:\ 161).

The disagreement between the model predictions and measured distributions for
the $\tau_0<1$ systems is more severe. None of the models is able to reproduce
the measured $b$ distribution. While the CHDM and tCDM models predict the
correct shape for the $\NHI$ distribution (both models give $\dks=0.06$ and
 $\Pks=0.2$), they predict far too great a number of optically thin
($\tau_0<1$) lines compared with the measured number of 316, as do all
the models except OCDM (CHDM:\ 395; \LCDMLns:\ 388; \LCDMHns:\ 413;
OCDM:\ 357; SCDM:\ 414; tCDM:\ 426). Thus it appears the discrepancy between
the model predictions and measured distributions arises primarily from the
optically thin systems, although there is still a pronounced tendency for the
models that agree best with the pixel flux distribution to predict too low
$b$-values even for the optically thick ($\tau_0>1$) absorbers.

We note that the measured median Doppler parameter for the optically thin
systems, $\bmed=28.6\pm1.3$\kms, is significantly higher than for the optically
thick systems, $\bmed=25.6\pm1.0$\kms. This is opposite the trend of a
decreasing envelope of $b$-values for optically thin systems found at lower
redshift (Hu \etal 1995; Kirkman \& Tytler 1997). It is also opposite the trend
found in the simulations. The model predictions for $\bmed$ among the optically
thick systems are for CHDM:\ 27.7\kms; \LCDMLns:\ 20.9\kms; \LCDMHns:\
23.1\kms; OCDM:\ 18.8\kms; SCDM:\ 20.4\kms; tCDM:\ 25.3\kms. Of these, only
the CHDM, \LCDMHns, and tCDM models are consistent with the measured value.
The predictions for $\bmed$ for the optically thin systems, however, are
substantially smaller:\ CHDM:\ 24.6\kms; \LCDMLns:\ 18.4\kms; \LCDMHns:\
19.6\kms; OCDM:\ 18.2\kms; SCDM:\ 18.5\kms; tCDM:\ 22.1\kms. Only the
CHDM model is able to match the high measured value within $3\sigma$. We
return to a discussion of this discrepancy in \S~\ref{sec:discussion_lp} below.

\begin{figure}
\begin{center}
\leavevmode \epsfxsize=3.3in \epsfbox{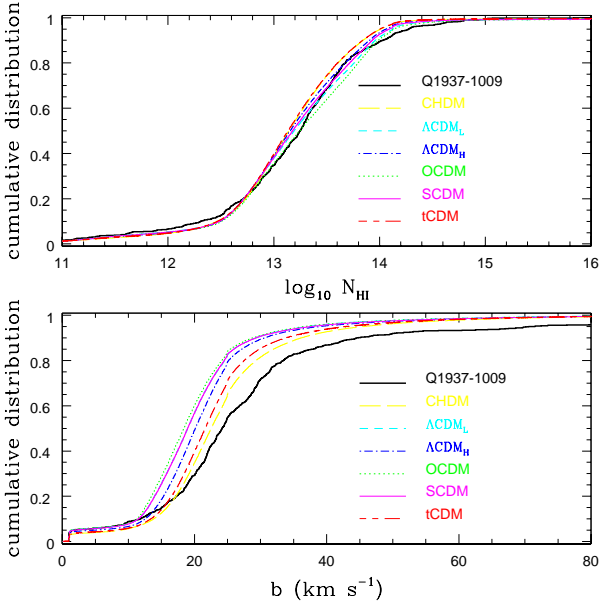}
\end{center}
\caption{Comparison between the measured absorption line parameter
cumulative distributions of Q1937$-$1009 and the model predictions using
\SPECFITfc, allowing for an offset in the continuum level.}

\label{fig:BT97_lpcd_off_spf}
\end{figure}

To test the sensitivity of the distributions of absorption line parameters to
the line-analysis method, we perform a second set of analyses using \SPECFIT.
The results are shown in Fig.~\ref{fig:BT97_lpcd_off_spf}. The adopted
threshold optical depth for finding lines is $\tau_{\rm min}=0.15$, although
smaller optical depth lines may result from the fits. (It would have been
possible to use a smaller value for $\tau_{\rm min}$ for most of the models;
the value adopted was required by the CHDM and tCDM models to avoid producing
an excessive number of lines. The same value
was adopted for the analysis of all the models for consistency.) The best fits
to the measured $\NHI$ distribution are provided by the \LCDML and
SCDM models, allowing for continuum offset corrections. The predicted CHDM and
tCDM $\NHI$ distributions do most poorly, both being significantly steeper than
the measured distribution. The KS test applied to the
\LCDML and SCDM results gives, respectively, $\dks=0.046$, $\Pks=0.16$ and
$\dks=0.056$, $\Pks=0.043$. The remaining models do significantly more poorly,
giving for CHDM:\ $\dks=0.086$, $\Pks=2\times10^{-4}$; \LCDMHns:\ $\dks=0.073$,
$\Pks=0.003$; OCDM:\ $\dks=0.073$, $\Pks=0.003$; tCDM:\ $\dks=0.092$,
$\Pks=6\times10^{-5}$. In contrast to the \AutoVP\ analysis, for which the CHDM
model provides a marginally acceptable fit, the $\NHI$ distribution derived for
CHDM using \SPECFIT\ provides a poor match to the data. By contrast, while the
\SPECFIT\ distributions for the \LCDML and SCDM models are acceptable, the
\AutoVP\ distributions for these models are strongly rejected.

As was found in the \AutoVP\ analysis, the models predict too low $b$-values:\
none of the models produces a $b$ distribution consistent with the measured
distribution. The measured median $b$-value is $24.2\pm1.8$\kms, where the
large error is due to the wide dispersion in $b$-values. All the model
predictions are smaller, but only that of OCDM by more than $3\sigma$.
The \SPECFIT\ analysis generally shows good consistency in the
predicted number of absorption features. The analysis of the spectrum of
Q1937$-$1009 yields 609 absorption features. The expected numbers of lines
predicted by the models are for CHDM:\ 658; \LCDMLns:\ 527;
\LCDMHns:\ 596; OCDM:\ 462; SCDM:\ 557, and tCDM:\ 661. Allowing for
Poisson fluctuations, all the model predictions are consistent with the
measured number except for \LCDML and OCDM, which are too low. Thus, although
the \LCDML model predicts the correct shape for the $\NHI$ distribution, it
predicts an incorrect normalization.

The differences in the \AutoVP\ and \SPECFIT\ results may be due in part to
the different relative emphasis the algorithms give to optically thin and
optically thick systems. Splitting the \SPECFIT\ results at $\tau_0=1$ shows
that while all the models predict a number of lines with $\tau_0>1$ consistent
(at the $3\sigma$ level) with the measured number of 178 (CHDM:\ 165;
\LCDMLns:\ 192; \LCDMHns:\ 185; OCDM:\ 189; SCDM:\ 194, and tCDM:\ 170),
only the CHDM, \LCDMH and tCDM models predict a number of optically
thin lines consistent with the measured number of 431. (These models predict
493, 412 and 491 lines, respectively.) The OCDM and \LCDML predictions lie over
$3\sigma$ too low, while the CHDM and tCDM predictions are marginally
consistent with the $3\sigma$ upper limit. This contrasts with the \AutoVP\
results, for which all the models predicted too many optically
thin systems. The two algorithms also yield differing levels of acceptability
of the shapes of the $\NHI$ distributions for the optically thin systems. While
the \AutoVP\ distributions agree well with the data for the CHDM and tCDM
models, the \SPECFIT\ results are inconsistent with the data for all the
models, except for the marginal case of tCDM ($\dks=0.09$, $\Pks=0.001$).

The results of \AutoVP\ and \SPECFIT\ agree more closely for the optically
thick systems, although still with some disagreement. While the
\AutoVP\ analysis yields acceptable $\NHI$ distributions for all the models
(although the results for the SCDM and tCDM models are marginal),
the \SPECFIT\ analysis yields distributions that are rejected for all the
models except CHDM ($\dks=0.08$, $\Pks=0.01$), although the \LCDML and OCDM
predictions are marginal (both give $\dks=0.14$ and $\Pks=0.001$). Thus the
acceptability of the predicted $\NHI$ distributions is somewhat sensitive to
the line-fitting algorithm used. This is due in part to the marginality of
the rejection levels:\ a small change in the distributions pushes the level of
agreement above or below the $3\sigma$ rejection threshold. Both algorithms
nevertheless agree that the CHDM model provides the best match to the measured
$\NHI$ distribution for the optically thick systems.

No significant difference is found in the median measured $b$-values
for the optically thin and thick systems using \SPECFIT. For $\tau_0<1$,
$\bmed=24.5\pm2.3$\kms; for $\tau_0>1$, $\bmed=24.1\pm2.2$\kms. Although all
the models predict lower values for the median $b$-value than measured, for
both $\tau_0<1$ and $\tau_0>1$ systems, the differences are not highly
significant given the large errors in the measured values. A stronger
statistical statement is made using the full sample. We note that the models
predict that the optically thin systems will have smaller median Doppler
parameters than the optically thick systems, but the differences are small,
amounting to only 1--2\kms.

Just as is found for the full sample, the predicted $b$ distributions for both
the $\tau_0<1$ and $\tau_0>1$ subsamples are rejected by the KS test for all
the models. The only exception is the marginal agreement found for the CHDM
model for the optically thin systems ($\dks=0.13$, $\Pks=0.004$). The \SPECFIT\
analysis consistently finds that the models predict too low $b$-values compared
with the data. This contrasts with the \AutoVP\ analysis, for which the
$b$ distribution predicted by tCDM for optically thick systems provides an
excellent match to the data (while all other model predictions are rejected).

It appears difficult to reach any definite conclusion regarding the
goodness-of-fit of the predicted Doppler parameter distributions to the data.
It is possible that \AutoVP\ and \SPECFIT\ cue off low optical depth
fluctuations differently. The main lesson is likely that there is a limit to
the physical interpretation of gas temperature or velocity dispersion that may
be attached to the fit Doppler parameters.

\subsection{HS~1946$+$7658}

Recent CMB experiments strongly favour a flat, or nearly flat, universe
(Netterfield \etal 2001; Pryke \etal 2001; Stompor \etal 2001).  From hereon,
we restrict the consideration of models to those currently most
viable:\ CHDM, the two \LCDM models, and tCDM. Although the tCDM model
is perhaps already excluded by the CMB data, it provides useful
comparisons with the predictions of the CHDM model and illustrates the
degree to which simulations of the \Lya forest may (or may not) be
used to discriminate between cosmological models based on the tests
considered in this paper.

\subsubsection{Flux distribution}

Over the redshift range $2.5<z<3.0$, the mean optical depth in the spectrum
of HS~1946$+$7658 is found to be $\bar\tau_\alpha=0.24$. The adopted optical
depth rescalings and conitnuum offset corrections
for these models are provided in Table~\ref{tab:KT97_rescale}.
No additional evolution in the UV background was required ($p=0$).

\begin{figure}
\begin{center}
\leavevmode \epsfxsize=3.3in \epsfbox{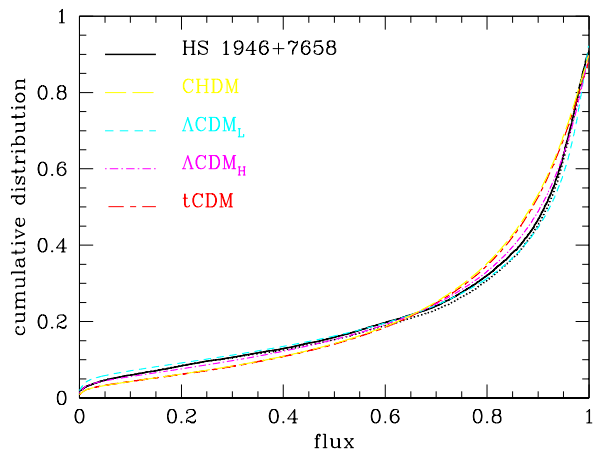}
\end{center}
\caption{Comparison between the measured flux distribution of HS~1946$+$7658
and the model predictions. A continuum offset correction
is applied to each model to
enforce matching the cumulative distributions at flux values near unity. The
dotted curve shows the measured cumulative flux distribution after removing
pixels contaminated by identified metal lines. The best matches are obtained
by the \LCDML and \LCDMH models, with predicted distributions that bracket the
measured one.}
\label{fig:KT97_fluxdist_off}
\end{figure}

The predicted and measured
distributions of flux per pixel (excluding bad pixels and the region
occupied by a Damped \Lya Absorber for the measured distribution), are shown in
Fig.~\ref{fig:KT97_fluxdist_off}. Also shown is the measured flux per pixel
distribution after removing pixels contaminated by metal lines identified by
Kirkman \& Tytler (1997). (We note that $\bar\tau_\alpha$ is negligibly
affected by the removal of these regions.) The best matches to the measured
metal-free distribution are provided by the \LCDML and \LCDMH models, with the
respective maximum cumulative distribution differences from that measured of
$\dks=0.047$ and $\dks=0.039$. Both models are strongly ruled out by the
formal KS test probability. Allowing for a reduction in the effective number of
pixels by a factor 3--5 (Appendix~\ref{app:tests}) still yields unacceptable
probabilities for the predictions. The two model distributions, however,
bracket the measured distribution, so it is reasonable to expect a \LCDM model
with an intermediate value of $\sigma_{\rm J}$ would provide an acceptable fit.

\subsubsection{Wavelet coefficient distributions}

\begin{figure}
\begin{center}
\leavevmode \epsfxsize=3.3in \epsfbox{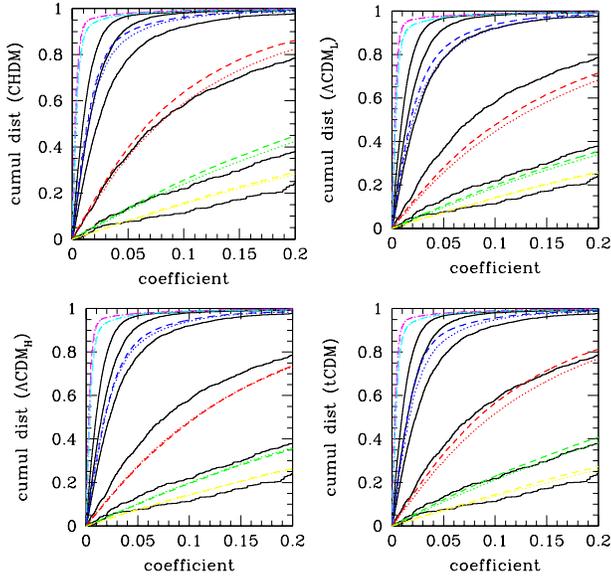}
\end{center}
\caption{Comparison between the measured wavelet coefficient cumulative
distributions of HS~1946$+$7658 (solid curves) and the model predictions
extrapolating about the simulation data at $z=3$ (dashed curves) and
interpolating between the simulation data at $z=2$ and $z=3$ (dotted curves),
allowing for continuum offset corrections in both. (For \LCDMHns, the dashed
curves correspond to interpolation between the simulation data at $z=2.7$ and
$z=3$.) The curves from left to right correspond to the velocity scales 2--4,
4--8, 8--16, 16--32, 32-64, and 64-128\kms. The best match for the velocity
scale 16--32\kms, corresponding to the median measured Doppler parameter
of the absorption features, is found for the CHDM and tCDM models. All the
models underpredict the amount of velocity structure on smaller scales.}
\label{fig:KT97_wc_off}
\end{figure}

The wavelet coefficient distributions are shown in
Fig.~\ref{fig:KT97_wc_off}. All the models poorly match the velocity
structure on scales of 8--16\kms and smaller, indicating the presence of
substantially more structure in the data on these scales than predicted by
the models. This structure reflects at least in part the presence of metal
lines.

Because the IGM temperature decreases with time in underdense regions, within
which the more abundant optically thin features arise, it is
unclear whether the best strategy for generating the distributions is by
interpolating beteen the simulation results at $z=2$ and $z=3$ or by
extrapolating the optical depths from the $z=3$ results alone, as the spectral
range analyzed lies nearer $z=3$ than $z=2$. The distributions for both
approaches are shown in Fig.~\ref{fig:KT97_wc_off}. The results of both
methods are qualitatively similar. (We note that the two approaches produce
negligible differences in the flux distributions.) Neither approach yields a
good match at low velocity scales. On the scale 16--32\kms, corresponding to
the median measured Doppler parameter (see below), acceptable agreement is
found for the CHDM results based only on interpolation ($\dks=0.042$,
$\Pks=0.026$), while agreement for tCDM is found only for extrapolating from
$z=3$ ($\dks=0.045$, $\Pks=0.015$). The agreement tends to improve for all the
models on larger velocity scales.

\subsubsection{Absorption line parameter distributions}

\begin{figure}
\begin{center}
\leavevmode \epsfxsize=3.3in \epsfbox{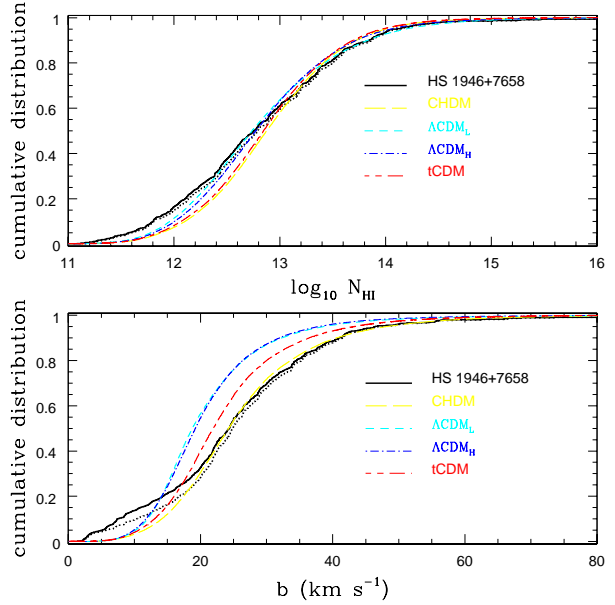}
\end{center}
\caption{Comparison between the measured absorption line parameter
distributions of HS~1946$+$7658 (solid curves) and the model predictions.
While good agreement is generally found for the $\NHI$ distributions above
$10^{13}$\cm2, all the models underpredict the number of lower column density
systems. The CHDM model best reproduces the measured $b$ distribution. The
agreements improve when metal absorption systems identified in the observed
spectrum are removed (dotted curves).}
\label{fig:KT97_lpd_off}
\end{figure}

\begin{figure}
\begin{center}
\leavevmode \epsfxsize=3.3in \epsfbox{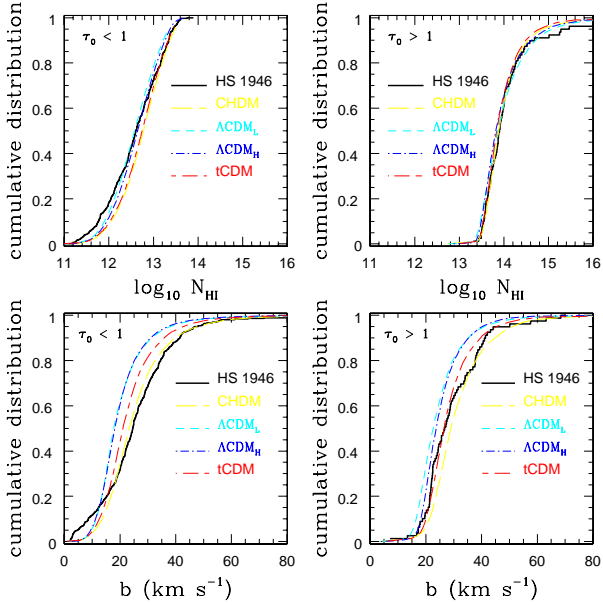}
\end{center}
\caption{Comparison between the measured absorption line parameter
distributions of HS~1946$+$7658 and the model predictions,
for systems with $\tau_0<1$ and $\tau_0>1$. The predicted distributions agree
better with the measured distributions for the optically thick systems
than for the optically thin.}
\label{fig:KT97_lpd_off_split}
\end{figure}

All the models underpredict the number of low \HI column density systems
and low $b$-value systems, as shown in Fig.~\ref{fig:KT97_lpd_off}.
It is possible that the low $\NHI$ and low $b$-value systems in the measured
spectrum are dominated by metal absorption features. Kirkman \& Tytler (1997)
identify a large number of metal lines in the redshift range analysed.
Unfortunately, because these lines are frequently blended with \Lya forest
systems, it is not possible to match them on a one-to-one basis with the lines
produced by \AutoVP. In an attempt to account for the metal absorbers,
we remove a narrow line ($b<15$\kms) from our list when it matches against a
metal absorber in the list of Kirkman \& Tytler (1997) to within a tolerance of
2\AA\ in the (observed) line-centre wavelength. This results in the rejection
of 28 systems. The resulting cumulative distributions, shown by the dotted
curves in Fig.~\ref{fig:KT97_lpd_off}, match the predicted more closely.
Formally, all the $\NHI$ distributions are acceptable at the $3\sigma$ level by
the KS test. The best matches are provided by the \LCDML and \LCDMH models,
which give for \LCDMLns:\ $\dks=0.063$, $\Pks=0.057$; and for
\LCDMHns:\ $\dks=0.064$, $\Pks=0.053$. Similar
results are obtained either by interpolating between the $z=2$ and $z=3$
simulation results, or extrapolating from the $z=3$ results alone. Only the
CHDM model yields a $b$ distribution consistent with the data. The KS test
gives $\dks=0.065$, $\Pks=0.048$, extrapolating from the $z=3$ simulation
results. An inconsistent distribution results when based on interpolation
between the $z=2$ and $z=3$ results because of the lower $b$-values at $z=2$.

As was the case for Q1937$-$1009, the models that provide the best fits the
flux distribution (\LCDML and \LCDMH) predict too small median Doppler
parameters. The measured median $b$ (after excluding metal lines) is
$24.9\pm0.8$\kms, while \LCDML and \LCDMH predict, respectively, 18.8\kms and
19.2\kms, both too small by over $7\sigma$.

The \AutoVP\ analysis of the spectrum of HS~1946$+$7658 yields 443
absorption lines. The predicted numbers for the various models are
for\ CHDM:\ 508; \LCDMLns:\ 539; \LCDMHns:\ 565; and tCDM:\ 539, all greatly
in excess of the measured number except for CHDM. The larger predicted numbers
are consistent with the general narrowness of the features.

As for Q1937$+$1009, we split the line samples into two subsamples, optically
thin ($\tau_0<1$) and optically thick ($\tau_0>1$) at line-centre. The
resulting $\NHI$ and $b$ distributions (with identified metal lines removed
as above) are shown in Fig.~\ref{fig:KT97_lpd_off_split}. Excellent agreement
is found between the model predictions for the $\NHI$ distribution and the
measured distribution for the optically thick systems. Acceptable agreement (at
the $3\sigma$ level) is found for the $b$ distributions as well, except for the
\LCDML model, which is marginally rejected by the KS test. The agreement for
the optically thin systems is much poorer. Only the \LCDM models yield
acceptable $\NHI$ distributions, while only the CHDM model predicts a
$b$ distribution consistent with the measured. The measured $b$ distribution
disagrees most strongly with the CHDM prediction at $b<15$\kms. Although we
match 28 metal lines with the list of Kirkman \& Tytler, they report 45 in the
redshift range of interest, which could potentially account for the difference.
We note that these CHDM results are based on extrapolating from the $z=3$
simulation results; the distribution derived by interpolating between the
$z=2$ and $z=3$ results yields unacceptably small $b$-values, as for the
other models.

The \AutoVP\ analysis of the spectrum of HS~1946$+$7658 yields 79 absorption
systems with $\tau_0>1$ and 364 with $\tau_0<1$. The predicted numbers of
optically thick features for the various models are for\ CHDM:\ 68;
\LCDMLns:\ 102; \LCDMHns:\ 92; and tCDM:\ 76, all consistent with the measured
number within Poisson fluctuations. The predicted numbers of optically thin
features for the models are for\ CHDM:\ 440; \LCDMLns:\ 437; \LCDMHns:\ 473;
and tCDM:\ 463. All the model predictions greatly exceed the number of measured
lines.

\subsection{Q0014$+$813, Q0302$-$003, Q0636$+$680, Q0956$+$122}

\subsubsection{Flux distribution}

The analysis redshift ranges for each of the Hu \etal (1995) QSO spectra are
given in Table~\ref{tab:qsos}. The mean optical depth found for each is:\
Q0014$+$813:\ $\bar\tau_\alpha=0.30$; Q0302$-$003:\ $\bar\tau_\alpha=0.31$;
Q0636$+$680:\ $\bar\tau_\alpha=0.29$; Q0956$+$122:\ $\bar\tau_\alpha=0.25$.
It was found unnecessary to include any amount of additional evolution in the
UV radiation background. The adopted optical depth rescalings and continuum
offsets are given in Table~\ref{tab:Hu95_rescale}.

\begin{figure}
\begin{center}
\leavevmode \epsfxsize=3.3in \epsfbox{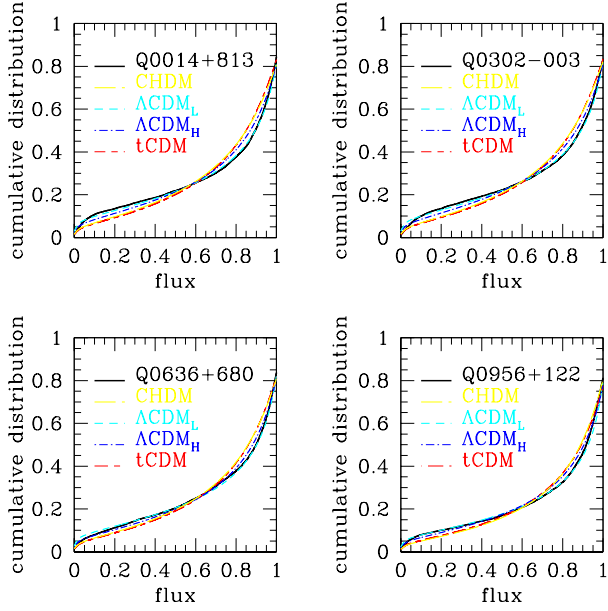}
\end{center}
\caption{Comparison between the measured flux distributions of the Hu \etal
(1995) QSO spectra and the model predictions. A continuum offset was applied to
each model to enforce matching the predicted and measured cumulative
distributions at flux values near unity. For all spectra except Q0956$+$122,
the \LCDML model provides the best match. For Q0956$+$122. the best match is
given by the \LCDMH model.}
\label{fig:Hu95_fluxdist_off}
\end{figure}

The distributions of flux per pixel are shown in
Fig.~\ref{fig:Hu95_fluxdist_off}. The best-fitting model is \LCDML in all
cases except Q0956$+$122, for which it is \LCDMHns. Formally, all the models
are rejected by the KS test. Allowing for a reduction in the effective number
of pixels by a factor of 3--5 due to correlations between neighbouring pixel
values yields acceptable agreement. The maximum differences between the
measured cumulative distributions and those predicted by the \LCDML model,
and the associated probabilities, are for Q0014$+$813:\ $\dks=0.031$,
$\Pks=0.01-0.09$; Q0302$-$003:\ $\dks=0.030$, $\Pks=0.02-0.11$; and
Q0636$+$690:\ $\dks=0.024$, $\Pks=0.09-0.31$. The \LCDMH model gives for
Q0956$+$122:\ $\dks=0.035$, $\Pks=0.003-0.04$. As was found for Q1937$-$1009
and HS~1946$+$7658 above, the \LCDML and \LCDMH model predictions tend to
bracket the measured distributions. The CHDM and tCDM models are
strongly rejected by all the QSO spectra, even allowing for possible
correlations in the pixel flux values.

\subsubsection{Wavelet coefficient distributions}

\begin{figure}
\begin{center}
\leavevmode \epsfxsize=3.3in \epsfbox{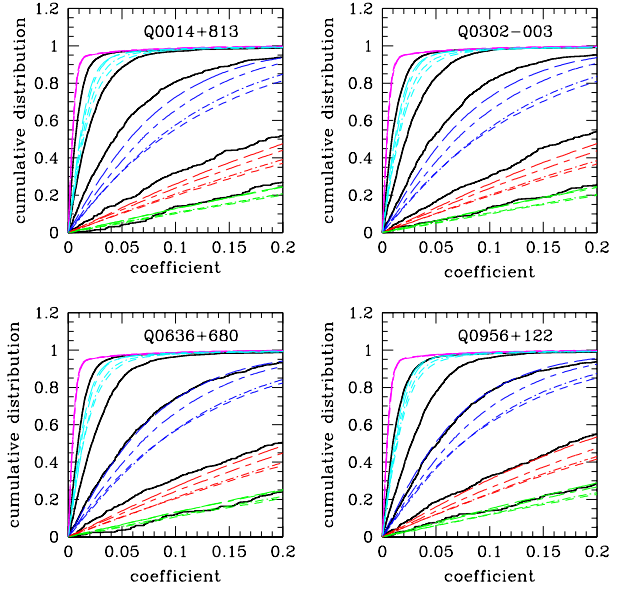}
\end{center}
\caption{Comparison between the measured wavelet coefficient cumulative
distributions of the Hu \etal (1995) QSO spectra (heavy solid curves) and
the model predictions, allowing for continuum offset corrections. The curves
from left to right correspond to the approximate velocity scales 4--8,
8--16, 16--32, 32--64, and 64--128\kms. Model predictions are shown
for CHDM (long-dash), \LCDML (short dash), \LCDMH (dot -- short dash), and tCDM
(short dash -- long dash). The best match for the velocity scale 16--32\kms,
corresponding to the median measured Doppler parameters, is found for the CHDM
model in all cases. All the models underpredict the amount of structure on
smaller velocity scales.}
\label{fig:Hu95_wc_off}
\end{figure}

As for HS~1946$+$7658, the predicted model distributions for the
wavelet coefficients are computed by
extrapolating about the simulation results at $z=3$, as this is closest
to the redshift ranges of the spectra. The values for $\alpha$ from
Table~\ref{tab:KT97_rescale} are used in the extrapolation.
The resulting wavelet coefficient
distributions are displayed in Fig.~\ref{fig:Hu95_wc_off}. The agreement
between the model predictions and the data is poor at the pixel scale
(4--8\kms). As was found in the previous QSO spectra, the data show a greater
amount of structure. The data continue to show more structure
on the 8--16\kms scale as well. Since the spectra resolve features on these
scales, the differences appear to be physical. At the next level (16--32\kms)
the trend reverses:\ the models predict more structure than found in the data.
The best agreement is consistently found for the CHDM model. The agreement,
however, is statistically acceptable only for Q0636$+$680 and Q0956$+$122,
according to the KS test. (For Q0636$+$680, $\dks=0.018$, $\Pks=0.88$, and
for Q0956$+$122, $\dks=0.035$, $\Pks=0.17$.) The level of agreement shows
significant scatter. The CHDM model predictions
at 16--32\kms agree well with the data for Q0636$+$680 and
Q0956$+$122, but not for the remaining two QSOs. The CHDM model performs
acceptably well on larger scales for all the QSOs.

\subsubsection{Absorption line parameter distributions}

\begin{figure}
\begin{center}
\leavevmode \epsfxsize=3.3in \epsfbox{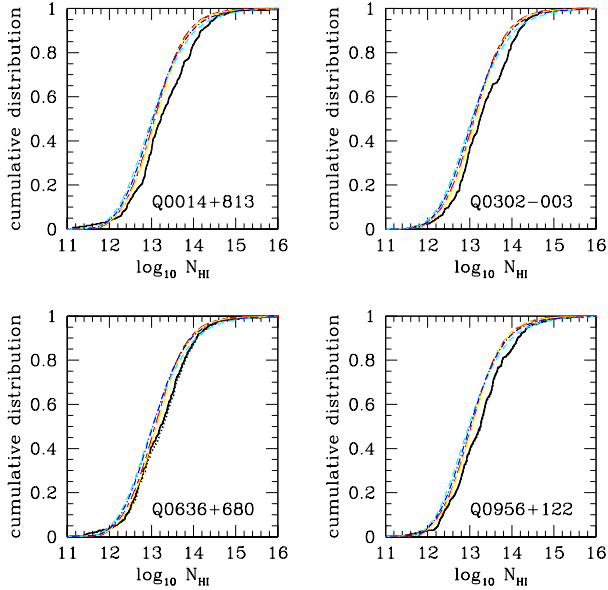}
\end{center}
\caption{Comparison between the \HI column density distributions measured
using the Hu \etal (1995) QSO spectra and the model predictions. The curves are
labelled as in Fig.~\ref{fig:Hu95_wc_off}. All the models predict a steeper
distribution than measured. The dotted curves show the distributions after the
removal of identified metal lines.}
\label{fig:Hu95_NHI_off}
\end{figure}

\begin{figure}
\begin{center}
\leavevmode \epsfxsize=3.3in \epsfbox{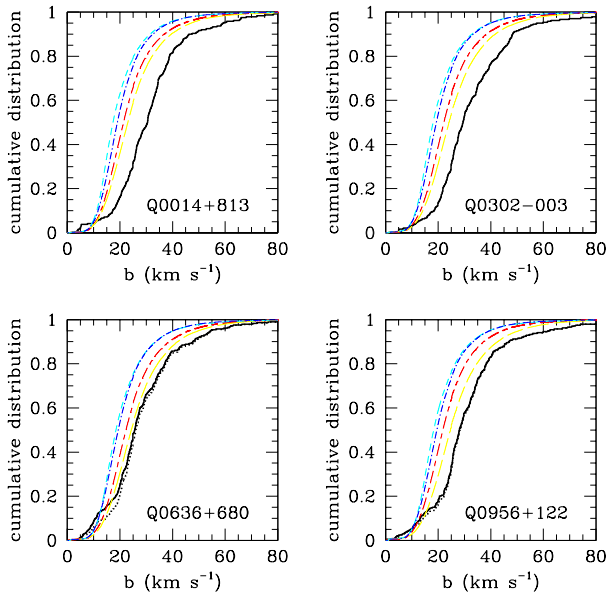}
\end{center}
\caption{Comparison between the Doppler parameter distributions measured
using the Hu \etal (1995) QSO spectra and the model predictions. 
The curves are labelled as in Fig.~\ref{fig:Hu95_wc_off}.
All the models predict a distribution peaking at lower values than measured.
The dotted curves show the distributions after the removal of identified metal
lines.}
\label{fig:Hu95_b_off}
\end{figure}

The \HI column density and Doppler parameter distributions derived by the
\AutoVP\ analyses of the QSO spectra of Hu \etal (1995) are shown in
Figs~\ref{fig:Hu95_NHI_off} and \ref{fig:Hu95_b_off} along with the model
predictions. All the models predict an \HI column density distribution and
Doppler parameter distribution that both peak at lower values than measured.
Of the four QSO spectra, the $\NHI$ distribution for only Q0636$+$680 is
well-represented by any of the models. The prediction of CHDM, the
best-fitting model, for the $\NHI$ distribution of Q0636$+$680 gives
$\dks=0.11$ and $\Pks=0.004$, while all other models are rejected. For all
the remaining QSO spectra, CHDM provides the best match to the measured $\NHI$
distribution as well, although the predicted distributions are rejected by
the KS test.

None of the models provides an acceptable match to the measured
$b$ distributions, although the CHDM model consistently comes closest. As found
for Q1937$-$1009 and HS~1946$+$7658, the model (\LCDML or \LCDMH) which best
agrees with the measured flux per pixel distribution produces Doppler
parameters that are too small.
The measured median Doppler parameters (after removing metal lines,
as discussed below), for Q0014$+$813, Q0302$-$003, Q0636$+$680, and Q0956$+$122
are, respectively, $30.5\pm1.2$\kms, $29.5\pm1.2$\kms, $26.1\pm1.1$\kms, and
$28.4\pm1.3$\kms. The predictions of \LCDML range between 17--18\kms for all
the QSOs, and for \LCDMH the prediction is 19\kms, all too small
by 35--75 per cent and by at least $6\sigma$.

The \AutoVP\ analysis of Q0636$+$680 produces 272 lines (after removing
identified metal lines). The CHDM model predicts 334 lines, greatly in
excess of the measured number. Similar or larger discrepancies between
the predicted and measured number of absorption lines are found for all
the QSO spectra and models, generally reflecting the narrowness of the
predicted absorption features compared with the measured.

We note that Hu \etal (1995) report a significantly larger number of narrow
absorption features in the spectrum of Q0636$+$680 than in any of the other
spectra, and that most of these may be identified with metal lines (13 lines
in the redshift range analysed here). They consequently exclude Q0636$+$680
from their statistics of the \Lya forest. We similarly find a larger number of
narrow lines ($b<10$\kms) in Q0636$+$680 compared with the other spectra. We
find that, based on the KS test, the flux and $\NHI$ distributions show no
anomalous behaviour compared with the other three spectra. In fact, these
distributions, along with the
$b$ distribution, agree better with the model predictions than the
distributions found for any of the other QSO spectra. The amount of metal line
contamination based on the tabulations in Hu \etal (1995) is small (5 per cent
of the lines in Q0636$+$680, 2 per cent in Q0956$+$122, no lines in
Q0014$+$813 and 1 line in Q0302$-$003). Removing the narrow lines in
the AutoVP analysis that most closely match the metal lines reported
in Hu \etal (1995) has little effect on the $\NHI$ and $b$ distributions,
as shown in Figs~\ref{fig:Hu95_NHI_off} and \ref{fig:Hu95_b_off}.
Although some of the narrow absorption features remaining may still be
unidentified metal lines, the similarity in the $\NHI$ distributions between
Q0636$+$680 and the other three QSOs suggests it is unlikely the contamination
is large, so that it would seem unlikely that unidentified metal lines
account for the large difference between the $b$ distribution of Q0636$+$680
and the other three QSOs. Moreover, there is no statistically
significant difference between the median Doppler parameters of the optically
thin ($\bmed=25.8\pm1.5$\kms) and optically thick ($\bmed=26.8\pm1.4$\kms)
systems in Q0636$+$680 (as is true for all the Hu \etal QSOs). If there were
still substantial metal contamination present, a significantly larger median
$b$-value for the optically thick systems may be expected. Because the median
$b$-values of the optically thin and thick systems are consistent for each of
the QSOs, the discrepancy between the predicted and measured Doppler parameters
found may not be attributed predominantly to the optically thin systems, unlike
the cases of Q1937$-$1009 and HS~1946$+$7658.

\subsection{J2233$-$606}

\subsubsection{Flux distribution}

The mean optical depth for J2233$-$606 is $\bar\tau_\alpha=0.16$ over the
analysis redshift interval $1.732<z<2.132$. The rescaling parameters used are
given in Table~\ref{tab:Hu95_rescale}. The simulation results at only $z=2$ are
used to extrapolate over the analysis redshift range, using the values for
$\alpha$ from Table~\ref{tab:KT97_rescale}. It was found unnecessary to
include any additional amount of evolution in the UV radiation background.

\begin{figure}
\begin{center}
\leavevmode \epsfxsize=3.3in \epsfbox{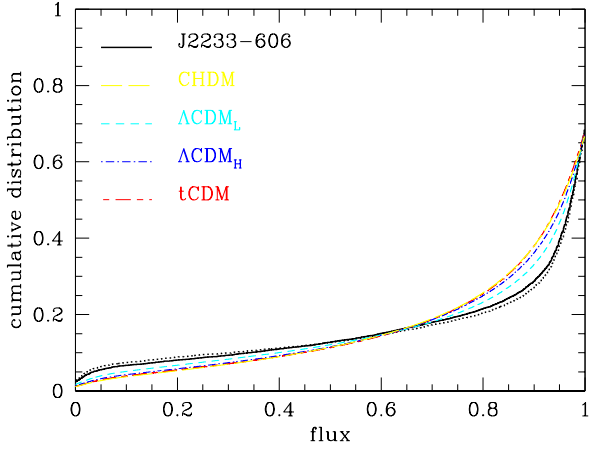}
\end{center}
\caption{Comparison between the measured flux per pixel distribution of
J2233$-$606 and the model predictions. A continuum offset was applied to each
model to enforce matching the cumulative distributions at flux values near
unity. None of the models recover the measured flux distribution, although
the \LCDML model is closest. The dotted line shows the measured flux
distribution after removing regions contaminated by identified metal lines.}
\label{fig:J22_fluxdist_off}
\end{figure}

The distributions of flux per pixel are shown in
Fig.~\ref{fig:J22_fluxdist_off}. All of the models differ greatly from
the data; the \LCDML model is closest. Cristiani \& D'Odorico (2000) have
identified a large number of metal absorption features in the spectrum. The
dotted line in Fig.~\ref{fig:J22_fluxdist_off} shows the flux distribution
after removing regions of the spectrum within 2\AA\ of an identified metal
line. The effect is to worsen the agreement with the models. The reason for
the poor agreement is unclear. At the low redshift of the QSO
($z_{\rm em}=2.238$), the fundamental mode across the simulation volume is
beginning to become non-linear. As a check, we have computed the flux per
pixel distributions by extrapolating the optical depths from the simulation
output at $z=3$ to the redshift range of the spectrum. Nearly
identical flux per pixel distributions result, so that the disagreement
appears not to be due to a lack of convergence in the simulation.

\subsubsection{Wavelet coefficient distributions}

\begin{figure}
\begin{center}
\leavevmode \epsfxsize=3.3in \epsfbox{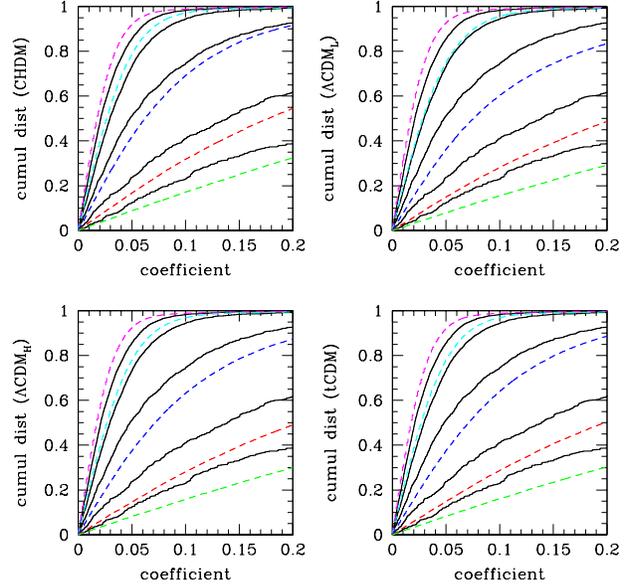}
\end{center}
\caption{Comparison between the measured wavelet coefficient cumulative
distributions of J2233$-$606 (solid curves) and the model predictions (dashed
curves), allowing for continuum offset corrections. The curves from left to
right correspond to the approximate velocity scales 4--7, 7--13,
13--27, 27--54, and 54--108\kms. The best match for the velocity scale
13--27\kms, corresponding to the median measured Doppler parameter, is found
for the CHDM model. All the models underpredict the amount
of velocity structure on smaller scales.}
\label{fig:J22_wc_off}
\end{figure}

The wavelet coefficient distributions are shown in Fig.~\ref{fig:J22_wc_off}.
All the models poorly match the velocity structure on scales of 7--13\kms and
smaller. The best agreement on the 13--27\kms scale, corresponding to the
measured median Doppler parameter (see below), is provided by the CHDM
model, although the match is rejected by the KS test ($\dks=0.09$,
$\Pks=10^{-9}$). The models recover the measured behaviour on larger
velocity scales only moderately well to poorly.

\subsubsection{Absorption line parameter distributions}

\begin{figure}
\begin{center}
\leavevmode \epsfxsize=3.3in \epsfbox{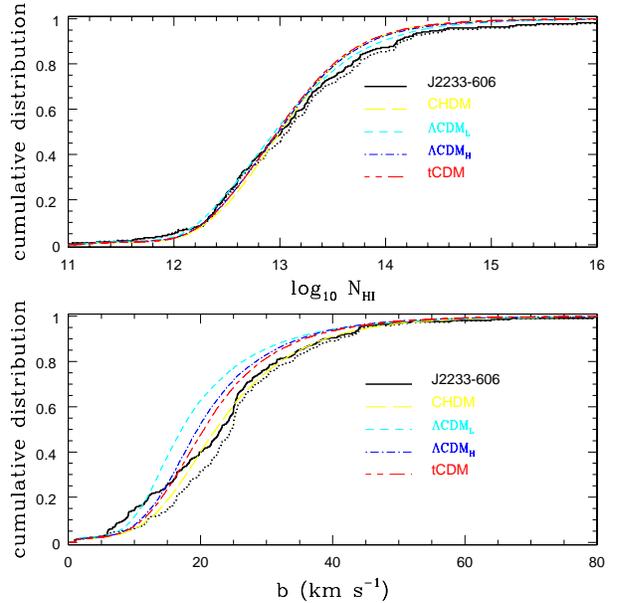}
\end{center}
\caption{Comparison between the measured absorption line parameter
distributions of J2233$-$606 and the model predictions. The dotted lines
show the measured distributions once identified metal lines have
been removed.}
\label{fig:J22_lpd_off}
\end{figure}

Despite the poor agreement with the measured flux distribution, all the models
well reproduce the $\NHI$ distribution, as shown in
Fig.~\ref{fig:J22_lpd_off}. (Using the line list of Cristiani \&
D'Odorico (2000), we match 28 of the narrow lines found to identified metal
lines and remove them.) The best agreement is given by the \LCDML model, with
$\dks=0.078$ and $\Pks=0.20$. The model predictions, however, disagree
in the number of lines. The predicted numbers are for CHDM:\ 306,
\LCDML:\ 304, \LCDMH:\ 312, and tCDM:\ 327, all greatly in excess of the
measured number of 192.

The model predictions for the $b$ distributions agree poorly with that
measured. The CHDM model agrees best, with $\dks=0.13$ and $\Pks=0.004$.
The predicted distributions of the remaining models are all strongly rejected.
As was found for all the previous QSO spectra, the model (\LCDML)
which provides the
best match to the flux per pixel distribution underpredicts the measured median
Doppler parameter. The measured median is $24.7\pm1.2$\kms, while \LCDML
predicts 17.2\kms, too small by $6\sigma$.

Splitting the sample into optically thin and optically thick systems tends to
worsen the agreement for the $\NHI$ predicted
distributions of the optically thick
absorbers to the point that they are only marginally acceptable (the best cases
are CHDM and \LCDMLns, both with $\dks=0.25$ and $\Pks=0.006$). The predicted
$\NHI$ distributions for the optically thin systems, however, all agree well
with the measured (optically thin) distribution. Only the CHDM and tCDM models
predict $b$ distributions consistent with the measured for the optically thick
absorbers (the tCDM prediction is marginal), while only the CHDM model predicts
a $b$ distribution consistent with the measured for the optically thin systems.
The agreement for CHDM tends to improve for both the $\tau_0<1$ and $\tau_0>1$
subsamples compared with the full sample. The number of optically thick lines
predicted by CHDM is 60, agreeing well with the measured number of 48. The
number of optically thin lines, however, agrees poorly:\ the predicted number
is 246, while only 144 are measured.
\section{Discussion}
\label{sec:discussion}

In this section we consider the possible systematics and limitations
of each of the tests employed, and discuss what may still be concluded
from the comparison between the models and the data. No single
cosmological model clearly emerges as the best based on the full set
of statistical tests. The measured distributions of flux per pixel are
the most accurately reproduced, with the agreement reaching high precision.
The agreement between the predicted and measured wavelet coefficients,
however, is substantially poorer. The data generally show considerably less
structure than the models predict on the velocity scale of approximately
15--30\kms. While the agreement often improves on larger velocity
scales, the best--fitting models tend to be those that provide poor fits
to the measured pixel flux distributions. The Voigt absorption line
decompositions of the spectra similarly display a conflict:\ most of the models
predict significantly narrower absorption features than measured, with the
best agreement obtained by those models which agree most poorly with the
measured flux per pixel distributions.

We show below that the shape of the flux per pixel distribution is sensitive to
the amount of power in the models on small scales. The absorption properties of
the spectra tend to be dominated by features that are marginally optically
thick at line-centre ($13.5<\log_{10}\NHI<14$), which are reproduced in the
simulations by structures with filamentary morphologies
(Miralda-Escud\'e \etal 1996; Zhang \etal 1998). The thicknesses of the
filaments correspond to the scale height expected for photoionized gas in
hydrostatic equilibrium within the gravitational potential of structures with
a moderate cosmological overdensity (Zhang \etal 1998). For small
density contrasts, this scale height is essentially the cosmological Jeans
length of the gas. We shall show that the flux distributions produced by
the absorbers vary systematically with the amount of power in the
linear power spectra on these scales. In the linear theory of the gravitational
growth of density fluctuations including baryons, the Fourier density modes
of the baryons are suppressed by the factor $1/[1+(k/k_{\rm J})^2]$, where
$k_{\rm J}=(4\pi G\rho a^2)^{1/2}/c_s$ is the (comoving) cosmological Jeans
wavenumber for a total cosmological density $\rho$ and baryonic sound speed
$c_s$, and $a$ is the expansion factor (Peebles 1984, 1993).
(The corresponding proper Jeans length is $L_J\equiv2\pi a/k_J=2\pi
(2/3)^{1/2}c_s/H(z)$, where $H(z)$ is the Hubble constant at redshift
$z$.) It is therefore useful to define the density fluctuations
$\sigma_{\rm J}$ filtered on the scale $k_{\rm J}$ according to
\begin{equation}
\sigma^2_{\rm J}=\int_0^\infty d\log k \frac{\Delta^2(k)}
{[1+(k/k_{\rm J})^2]^2},
\end{equation}
where $\Delta^2(k)$ is the dimensionless power spectrum $k^3P(k)/2\pi^2$
at the scale $k$. For simplicity, we define $\sigma_{\rm J}$ at $z=3$,
choosing $c_s$ to be the isothermal sound speed 16.8\kms for fully ionized gas
at a temperature $2\times10^4\kel$, the estimated temperature for gas with an
overdensity near unity at $z=3$ (McDonald \etal 2000a; Ricotti, Gnedin \&
Shull 2000; Schaye \etal 2000).

\subsection{Flux distribution}

The simulations are able to reproduce the measured flux per pixel distributions
to extremely high accuracy. The best case is given by the SCDM model for
Q1937$-$1009. The maximum difference between the measured and predicted
cumulative flux distributions is $\dks=0.016$, close to the level of
precision of the simulation (see Appendix~\ref{app:tests}). 
For HS~1946$+$7658, the \LCDMH model predicted distribution agrees with the
measured to $\dks<0.04$.  Similar levels of agreement are found for the other
QSO spectra examined, except for J2233$-$606 which appears anomalous.

Despite the close agreement between the predicted and measured flux
distributions, the formal statistical acceptability of the matches, as
given by the KS test, is generally poor. This is in part due to the large
number of pixels involved (approximately $10^4$), which permits a stringent
comparison to be made. It is, however, also due to deviations from the
theoretical KS probability distribution of $\dks$. In practice, the probability
distribution is broader than the theoretical, which is
likely a consequence of correlations between neighbouring pixel flux values
(see Appendix~\ref{app:tests}). Allowing for the correlations to decrease
the effective number of degrees-of-freedom yields acceptable agreement for
the predicted distributions of either the \LCDML or \LCDMH model (or both)
for all the QSOs (except HS~1946$+$7648, although the predicted distributions
of the two \LCDM models bracket the measured distribution, and J2233$+$606),
while the predictions of the CHDM and
tCDM models are consistently rejected. We note that Rauch \etal (1997) and
McDonald \etal (2000b) also found very close agreement between the predicted
flux distribution of a \LCDM model similar to our \LCDMH model and the measured
flux distribution in a set of QSO spectra observed with the Keck HIRES. In
contrast to our results, however, Rauch \etal found the predicted flux
distribution of an SCDM model nearly identical to ours to agree less well with
the measured.

The flux distributions are found to vary systematically with $\sigma_{\rm J}$,
the density fluctuations on the scale of the cosmological Jeans length. As
shown in Fig.~\ref{fig:BT97_fluxdist_off}, as $\sigma_{\rm J}$ increases, the
distributions flatten (as measured between flux values of 0.2 and 0.8).
This is similar to the trend found by M00 of increasing width of the
probability distribution of $\log\tau$ with increasing small-scale power. The
best-fitting predicted flux distribution for Q1937$-$1009 is provided by the
SCDM model, for which $\sigma_{\rm J}=1.6$. Although the predicted flux
distributions of the \LCDML and \LCDMH models agree less well, their cumulative
flux distributions bracket the measured distribution, as is found for all the
QSO spectra except J2233$+$606. The values of
$\sigma_{\rm J}$ for these models (1.7 and 1.3, respectively), similarly
bracket the value of $\sigma_{\rm J}$ for the SCDM model, so that it is
reasonable to expect a \LCDM model with an intermediate value of
$\sigma_{\rm J}$ would provide a better fit.

The success of the \LCDM models in predicting pixel flux distributions that
bracket the measured distributions suggests a strong bound on the small scale
power of the primordial power spectrum at $z=3$ of $1.3<\sigma_{\rm J}<1.7$
(with greater than $3\sigma$ certainty). This result combined with limits on
$\sigma_{8h^{-1}}$ from cluster abundances (Pierpaoli \etal 2001), on the
primordial power spectrum shape parameter ($0.19<\Gamma_{\rm PS}<0.37$;
Eisenstein \& Zaldarriaga 2000) and its amplitude (Bunn \& White 1997) and
spectral index ($0.9<n<1.1$; Netterfield \etal 2001; Pryke \etal 2001;
Stompor \etal 2001), imposes the limit on the cosmological matter density
$0.26<\Omega_M<0.43$ for a flat universe ($\Omega_\Lambda=1-\Omega_M$), and
assuming $\Omega_bh^2=0.020$ (O'Meara \etal 2000), $0.45<h<0.85$, and
no massive neutrino species. These restrictions ease somewhat to
$0.26<\Omega_M<0.68$ if up to two massive neutrino species are allowed
accounting for a total mass fraction $\Omega_\nu\le0.2$, and the $3\sigma$
limits of Pierpaoli \etal (2001) are permitted.
These limits are in good agreement with those found
elsewhere (eg, Netterfield \etal 2001; Pryke \etal 2001; Stompor \etal 2001).
Adopting $\Gamma_{\rm PS}=0.15$ in accordance with Croft \etal (2000), based
on the flux power spectrum $P_F(k)$ measured in several \Lya forest spectra,
gives (for $1.3<\sigma_J<1.7$) the limits $0.23<\Omega_M<0.45$
($\Omega_\Lambda=1-\Omega_M$, $\Omega_\nu=0$), in satisfactory
agreement with the limits of Croft \etal (2000).

\subsection{Evolution of $\Gamma$}

Since the gas is in photoionization equilibrium,
the optical depth in \Lya is predicted by the models only to within the factor
$b_{\rm ion}$ (\S~\ref{sec:results}). The re-normalizations of
the optical depths required for a given model to match the measured mean
optical depth in \Lya of a QSO spectrum provides an estimate of the value of
$\Gamma$ required.

\begin{figure}
\begin{center}
\leavevmode \epsfxsize=3.3in \epsfbox{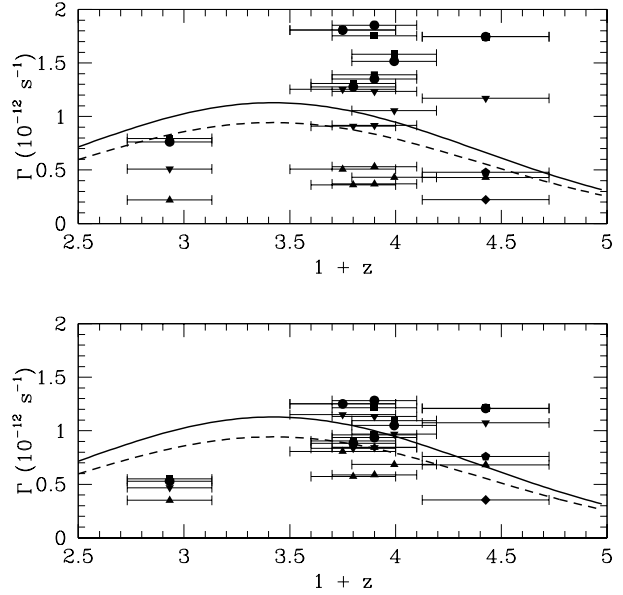}
\end{center}
\caption{Required \HI photoionization rate $\Gamma$ required for the models to
match the measured mean optical depths, for CHDM (circles), \LCDML (triangles),
\LCDMH (inverted triangles), tCDM (squares), OCDM (diamonds), and
SCDM (pentagons). The upper panel shows $\Gamma$ for the value of $\Omega_b$
assumed in the model. The lower panel shows $\Gamma$ assuming a fixed value
of $\Omega_bh^2=0.020$ for all the models. The predicted values are close
to the estimated $\Gamma$ from QSO sources alone for an intrinsic QSO spectral
index of $\alpha_Q=1.5$ (solid curve) and $\alpha_Q=1.8$ (dashed curve).}
\label{fig:Gamma}
\end{figure}

In Figure~\ref{fig:Gamma} we show the required values of $\Gamma$ for the
models to match the measured mean optical depth $\bar\tau_\alpha$ in each
QSO spectrum. The width of a given horizontal bar indicates the redshift range
analysed in the QSO spectrum, and the corresponding point marks the required
value of $\Gamma$ at the central redshift. Two estimates are made. For the
first, we show the values of
$\Gamma$ assuming the value for $\Omega_b$ adopted in each model. For the
second, we show the values of $\Gamma$ assuming $\Omega_bh^2=0.020$ (O'Meara
\etal 2000) after rescaling the optical depths by $b_{\rm ion}$.
The scatter is considerably reduced in the latter case. Also shown are
the theoretical predictions for $\Gamma$ due to QSO sources alone, assuming an
intrinsic QSO spectral index of $\alpha_Q=1.5$ or 1.8 (Haardt \& Madau 1996).
The required values of $\Gamma$ are found to lie within a factor of 2 of
the predicted rates for $\Omega_bh^2=0.020$, suggesting little, if any, sources
additional to QSOs are required, or are even permitted without destroying the
agreement between the models and the measured spectra. We note, however, that
the rate of evolution of $\Gamma$ necessary to match the measurements of the
mean optical depth over the redshift range probed by the spectrum of
Q1937--1009 ($3.1<z<3.7$) was found to be significantly slower than given by
Haardt \& Madau (1996). The difference may be due to the uncertainty in the
QSO counts at these high redshifts.

The values we obtain for $\Gamma$ for the \LCDMH model at $z=3$ is about
twice that reported by McDonald \etal (2000b) for a very similar model. (Here
and below, we use $b_{\rm ion}$ to re-normalize the reported values of $\Gamma$
to $\Omega b^2=0.020$.) The difference is likely due primarily to the larger
value for $\bar\tau_\alpha$ they find. Our results are in closer agreement with
those of Rauch \etal (1997) in the redshift intervals $2.5<z<3.5$ and
$3.5<z<4.5$, although our values for $\Gamma$ still lie somewhat higher. By
contrast, our value for $\Gamma$ for the SCDM model in the interval $3.1<z<3.7$
is about a factor of 2 below that found by Rauch \etal for SCDM in the range
$3.5<z<4.5$. The difference may in part stem from an overestimate of
$\bar\tau_\alpha$ by the SCDM model simulation used by Rauch \etal (that of
Croft \etal 1997), as was demonstrated by Theuns \etal (1998b). The value for
$\Gamma$ reported by Weinberg \etal (1997) at $z=3$ is just above the spread
of values we find for our various models.

\subsection{Wavelet coefficient distributions}

The wavelet coefficient distributions predicted by the models show considerable
disagreement with the measured distributions. The deviations are greatest on
the velocity scale of approximately 15--30\kms, with the models generally
predicting a much greater amount of structure than is found in the data.
The closest agreement is found for the CHDM model on this scale for all the
QSO spectra. The CHDM model predicts distributions on this scale consistent
with the data for Q1937$-$1009 and HS~1946$+$7658. The level of agreement,
however, is inconsistent among the Hu \etal (1995) QSOs on the 16--32\kms
scale. While good agreement is found for Q0636$+$680 and Q0956$+$122, the
CHDM predictions are strongly rejected by the KS test for Q0014$+$813 and
Q0302$-$003.

To investigate the reasons for the inconsistency in the level of agreement on
the 16--32\kms scale of a given model among the various Hu \etal QSO spectra,
we compare the measured wavelet distributions of these spectra. A comparison
of the distributions of Q0302$-$003 and Q0636$+$680 shows that they
are only marginally consistent on the 16--32\kms scale ($\dks=0.08$,
$\Pks=0.005$) (while consistent on the other scales examined). A comparison
between Q0302$-$003 and Q0956$+$122, however, shows the distributions
to be consistent on the scale of 16--32\kms and higher, but not on smaller
scales. The data show statistically significant inhomogeneity. For
this reason, we have chosen to treat the analyses of the QSOs separately.
(We note that the flux per pixel distributions of the four Hu \etal QSO spectra
show no significant differences over the analysis redshifts, and so in
principle could have been treated together. For uniformity of presentation,
we compare the numerical results with the results from each spectrum
individually as well.)

The level of agreement between the models and the data
is not much improved on the next smaller scale of approximately
8--16\kms. While the predicted distributions of coefficients for Q1937$-$1009
tend to scatter either above or below the measured, the predicted coefficients
are systematically too small on this scale for HS~1946$+$7658. The additional
measured structure in the latter spectrum may reflect
the large number of metal absorption lines identified
(Kirkman \& Tytler 1997), which will give rise to features on this scale.

None of the models reproduce the light fluctuations on the pixel scale, as
determined by the lowest level wavelet coefficients. It may be
that the disagreement results from limitations of the simulations. For an
expansion velocity difference across the simulation box at $z=3.5$ of
$\sim1000$\kms and 256 grid zones along a side, the characteristic velocity
resolution is $\sim4$\kms, so that it may be difficult to reproduce
fluctuations in the spectra on these scales.

Although the wavelet analysis provides a quick and precise means of comparing
the model predictions with the data, it has the disadvantage of not directly
relating any discrepancies to physical quantities. Larger (or smaller) measured
coefficients than predicted at a given velocity scale will arise because fewer
(or more) absorption systems are predicted by the model on this scale, but
the wavelet coefficients alone cannot determine whether this is a consequence
of a mismatch of the density fluctuation spectrum, the assumed cosmological
model, the presence of contaminating lines (metal systems), or the thermal
properties of the gas. To examine these issues, we also perform
a Voigt absorption line analysis.

\subsection{Absorption line parameter distributions}
\label{sec:discussion_lp}

The shapes of the predicted and measured \HI column density distributions tend
to accord well, though not perfectly. The \AutoVP\ analysis of Q1937$-$1009
yields agreement with the measured $\NHI$ distribution only for the CHDM model,
and even this is marginal. The analysis of HS~1946$+$7658 (after removing
identified metal lines) favours instead the \LCDM models,
although marginal agreement is again found for CHDM and tCDM. The predicted
and measured $\NHI$ distributions for all the Hu \etal (1995) spectra
tend not to agree at an acceptable level, though the predictions of CHDM are
favoured. The agreement with the $\NHI$ distribution measured in J2233$-$606,
however, is acceptable for all the models. The level of agreement is somewhat
sensitive to the analysis method:\ better agreement is found for Q1937$-$1009
between the predicted and measured $\NHI$ distributions using \SPECFIT\ than
\AutoVP, though still at a marginal to poor level, except for \LCDML for which
the agreement is good. Of greater concern is the number of absorption systems
found:\ both the \AutoVP\ and \SPECFIT\ analyses predict numbers of features
inconsistent with the numbers measured.

Splitting the absorption line samples into subsamples optically thin and thick
at the \Lya line-centre reveals that the discrepancies originate primarily from
the optically thin systems. Acceptable agreement is found for the
optically thick systems between the predicted and measured $\NHI$ distributions
for Q1937$-$1009 for all the models using \AutoVP. (The agreement tends to
worsen for \SPECFIT, however, again suggesting that caution should
be exercised in judging the viability of a model on the basis of the $\NHI$
distribution alone.) The predicted and measured distributions for
HS~1946$+$7658 for the optically thick systems
coincide extremely closely for all the models examined, and acceptable
agreement is again obtained for J2233$-$606 (though at a more marginal
level). There is generally acceptable agreement between the predicted and
measured numbers of optically thick absorbers as well.

The measured and predicted $\NHI$ distributions for the optically thin systems
match much more poorly. While excellent agreement is found between the
predictions of the CHDM and tCDM models and the measured distribution in
Q1937$-$1009 using \AutoVP, all the other models are strongly rejected. (The
\SPECFIT\ analysis results in no acceptable models, though tCDM is marginal.)
Poorer agreement is found for HS~1946$+$7658 among the optically thin
systems than the optically thick as well, although the \LCDM models
are still acceptable. For J2233$-$606, the agreement with CHDM and tCDM is
excellent, while the \LCDM models are marginal.

We note that our results are somewhat at variance with those of Gnedin (1998),
who uses a pseudo-hydrodynamical numerical technique to test a variety of
cosmological models based on the predicted \HI column density distributions.
Gnedin finds acceptable agreement between the predictions of several models
for the shape of the $\NHI$ distribution with the tabulated distribution of
Hu \etal (1995). By contrast, we find none of our models recovers the Hu \etal
distribution, including our \LCDM models which are very similar to models
Gnedin finds acceptable. Based on the results of several models, Gnedin places
$2\sigma$ and $3\sigma$ constraints on the size of density fluctuations for
wavenumbers twice the Jeans wavenumber. Using his conversion formula to
$\sigma_{34}$, the density fluctuations gaussian smoothed on the scale
$k_{34}=34\Omega_M^{1/2}h\,{\rm Mpc}^{-1}$, where $\Omega_M$ is the total
matter density parameter, and the ``effective equation of state'' parameters
reported by Ricotti \etal (2000), these limits become ($z=2.85$)
$1.4<\sigma_{34}<2.6$ ($2\sigma$) and $1.3<\sigma_{34}<3.1$ ($3\sigma$).
These rule out our CHDM model ($\sigma_{34}=1.4$) at the $2\sigma$ level
and our tCDM model ($\sigma_{34}=1.1$) at the $3\sigma$ level, but not the
others. Yet we find that the shapes of the $\NHI$ distributions predicted by
CHDM and tCDM for Q1937$-$1009 (restricted to the optically thick systems,
corresponding approximately to the column density range used by Gnedin in the
same redshift range), are consistent with the measured (optically thick)
distribution, with the CHDM model providing the best match of all the models.
These two models also fare no worse than the others in comparison with the
Hu \etal (1995) data, and often provide the best (though usually unacceptable)
match to the measured $\NHI$ distribution. Whether the disagreement between
our findings and Gnedin's is due to differences in the absorption line analysis
approaches or the simulation methods (Meiksin \& White 2001) is unclear.

A much larger discrepancy is found between the measured Doppler parameter
distributions and the distributions predicted by the simulations. The median
measured Doppler parameters generally exceed the
predicted by as much as 30--60 per cent. The best agreement is consistently
found for the predictions of the CHDM model, although this model provides a
poor match to the measured flux per pixel distributions.
The next best match to the measured $b$ distributions is provided by tCDM,
which also predicts flux per pixel distributions in disagreement
with the data. On the other hand, the models that match the measured pixel flux
distributions predict Doppler parameters that are much too small. There appears
no simple remedy to this conflict.

Splitting the sample into optically thick and optically thin absorbers at the
\Lya line-centre shows that the discrepancy in the Doppler parameters is
sensitive to the line-centre optical depth $\tau_0$.
The discrepancy in the measured and predicted Doppler parameters has been
recognized for optically thick systems for \LCDMns, OCDM, and SCDM models
(Theuns \etal 1998b; Bryan \etal 1999; Theuns \etal 1999; Bryan \& Machacek
2000; M00). The results of M00, however, suggest that the discrepancy may be
model-dependent, as the tCDM model was found there to provide reasonably good
agreement. Here we find that the tCDM model provides an excellent match to the
measured $b$ distributions for the optically thick absorbers in Q1937$-$1009
and HS~1946$+$7658, while the $b$ distributions predicted (for $\tau_0>1$) by
the CHDM and \LCDMH models are consistent with that of HS~1946$+$7658, though
not of Q1937$-$1009.

None of the models matches the distributions measured in the Hu \etal (1995)
QSO spectra, although the CHDM model comes close for Q0636$+$680. This latter
point is of particular concern, since previous model comparisons have been
based on the Hu \etal QSO sample, supplemented by several other QSO spectra
(Kim \etal 1997), but with Q0636$+$680 {\it excluded}. The reason for excluding
it was the relatively large number of absorption features found, as well as the
large number of metal systems identified in the spectrum:\ it appeared that
the spectrum may be excessively contaminated by metal lines. We perform a
comparison between the \AutoVP\ results for Q0636$+$680 and the combined other
three QSOs of Hu \etal over the common redshift range $2.7<z<3.0$
(with identified metal lines removed), and confirm a statistically significant
difference in the $b$ distributions using the KS test ($\dks=0.16$,
$\Pks=6\times10^{-4}$). The $\NHI$ distributions, however, agree extremely
well, which may not have been expected if a large amount of metal contamination
was still present in the Q0636$+$680 line list. (The distributions of restframe
equivalent widths and line-centre optical depths also agree extremely well.)
The larger number of lines in Q0636$+$680 is expected given the relative
narrowness of the features (in order to recover the measured value of
$\bar\tau_\alpha$). Since all the spectra probe large cosmological
distances ($\Delta z/(1+z)\approx0.1$), sample variance would appear an
unlikely explanation for the discrepancy in $b$-values. The smaller median
$b$-value measured in Q0636$+$680 is also in better agreement with the median
$b$-value measured in HS~1946$+$7658, at only a slightly smaller redshift. We
conclude that there is little justification for preferring the results of the
other Hu \etal QSO spectra over those of Q0636$+$680. Adopting the tabulated
results of Hu \etal (1995) or Kim \etal (1997) as the fiducial
$b$ distribution for comparison with the simulation predictions may
exaggerate the true level of disagreement, if any.

More pronounced disagreement in the Doppler parameters is found for the
optically thin absorbers. Only the CHDM model is able to produce a distribution
consistent with any of those measured (only for HS~1945$+$7658 and J2233$-$606,
although the \SPECFIT\ results of the CHDM model for Q1937$-$1009 are
marginal). The discrepancy in the numbers of absorption systems is least for
the CHDM model. This may be attributed to the larger $b$-values:\ fewer
broader lines are required to produce the same mean optical depth
$\bar\tau_\alpha$. A similar explanation accounts for the better agreement
in the wavelet coefficient distributions compared with the other models on the
scale of 16--32\kms:\ fewer features on this scale are produced by the CHDM
model than by the other models, resulting in a smaller proportion of
large coefficients and so a better match to the measured distributions.

Possibly a compromise cosmological model may be found which fits all the
distributions. Increasing the baryon density will increase the temperature
of the moderate density gas and hence the Doppler parameters (Meiksin 1994;
Theuns \etal 1998b). For instance, doubling the baryon density for \LCDML from
$\Omega_bh^2=0.015$ to 0.03 would increase the gas temperature by 40 per cent,
using the temperature-baryon density scaling of
Zhang \etal (1998). This may increase the median Doppler parameter by as much
as 10--20 per cent, depending on the contribution of peculiar velocity
broadening to the line widths (Zhang \etal 1998). Although such a high baryon
density contradicts the deuterium abundance determinations of O'Meara \etal
(2000) within the context of standard Big Bang Nucleosynthesis, it is
consistent with reported CMB measurements of the acoustic peaks
(Netterfield \etal 2001; Pryke \etal 2001; Stompor \etal 2001).

There is an issue of convergence of the simulation results,
particularly for the $b$ distribution. Theuns \etal (1998b) and
Bryan \etal (1999) show that the $b$-values tend to decrease with
increasing spatial resolution of the simulations, although they were able
to explore the trend only for a limited range of box sizes.
Meiksin \& White (2001) expanded the box size and resolution ranges by
performing pseudo-hydrodynamical $N$-body simulations of the \Lya forest. They
found that convergence was particularly difficult to achieve in the
distribution of wavelet coefficients on the scale of $16-32$\kms and
in the $b$ distribution. The results of their Particle-Mesh simulations
suggest that the cumulative distributions may be determined only up to an
absolute precision of $0.05-0.1$, and the convergence does not always scale
monotonically with resolution. The convergence behaviour, however, was
found to improve when a pseudo-pressure force was added, and so may be better
for a full hydrodynamics computation as well. They were still unable to reach
convergence in the cumulative $b$ distribution to better than an absolute
precision of about 0.05 even in the pseudo-pressure force case, and
allowing for as many as $512^3$ grid zones to compute the gravity, a
greater number than used in the simulations here. The question of convergence
of the full hydrodynamics simulations to an absolute precision of better than
0.05 in the cumulative $b$ distribution we consider as yet unresolved. The
general tendency, however, is to reduce the Doppler parameters at increased
resolution, so that the discrepancy between the measured and predicted Doppler
parameters is unlikely to result from a lack of convergence in the simulations.

\begin{figure}
\begin{center}
\leavevmode \epsfxsize=3.3in \epsfbox{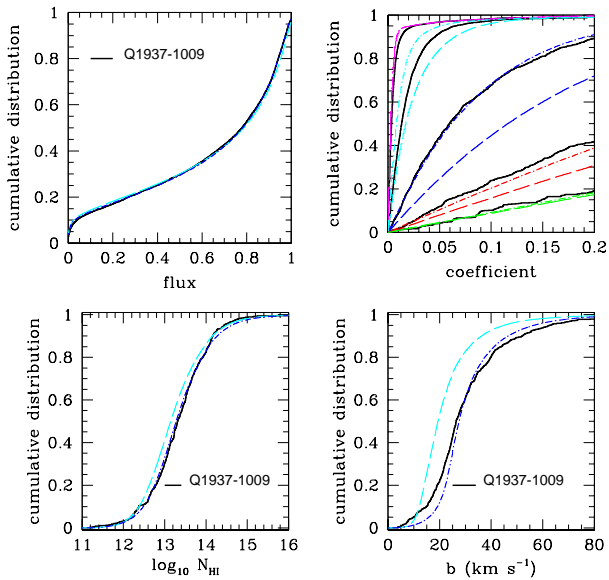}
\end{center}
\caption{The effect of additional heat input due to late \HeII reionization
on the predictions of the \LCDML model for Q1937$-$1009. Shown are the
cumulative distributions of the flux per pixel (top left panel), wavelet
coefficients (top right panel; labelled as in Fig.~\ref{fig:BT97_wc_off}),
\HI column density (bottom left panel) and Doppler parameter (bottom right
panel). The effect of late reionization is mimicked by broadening the
simulation optical depths by an additional 12\kms (dot-dashed lines),
corresponding to the expected increase in temperature due to the onset of
sudden \HeII reionization. The original results without the added broadening
are also shown (dashed curves), as well as the direct measurements from the
spectrum of Q1937$-$1009 (solid curves). The extra broadening has very little
effect on the flux per pixel distribution, but pronounced effects on the
wavelet coefficients and fit absorption line parameters.}
\label{fig:BT97_broadened}
\end{figure}

Another possibility is that there is missing physics in the simulations.
Late \HeII reionization will produce an increase in the temperature of the
gas (Meiksin 1994; Miralda-Escud\'e \& Rees 1994; Haehnelt \& Steinmetz 1998;
Abel \& Haehnelt 1999). The possibility of late \HeII reionization, at a
redshift of $z_{\rm HeII}=3-3.5$, has been suggested in a variety of
contexts: on the basis
of intergalactic \HeII optical depth measurements compared with \HI
(Madau \& Meiksin 1994; Reimers \etal 1997); an abrupt change in the ionization
parameter of metal lines at these redshifts (Songaila \& Cowie 1996, but this
has been disputed by Boksenberg, Sargent, \& Rauch 1998); and determinations of
an ``effective equation of state'' of the IGM (McDonald \etal 2000a;
Ricotti \etal 2000; Schaye \etal 2000). Since the optically thin absorption
originates preferentially in
underdense regions (Zhang \etal 1998), these systems will not achieve thermal
balance between photoionization heating and radiative cooling, and so will more
strongly manifest the thermal effects of late \HeII ionization (Meiksin 1994).
Since the associated heating will occur while the gas is out of
ionization equilibrium, the temperature will be boosted by as much as a factor
of 2 over the equilibrium case (Meiksin 1994). The low density optically thin
absorbers will thereafter cool largely through adiabatic expansion,
while the temperature of the higher density gas
will relax from its post-photoionization level to a lower value as a result of
radiative losses.

We may estimate the change in temperature immediately following \HeII
photoionization as follows. Prior to \HeII ionization, the total number density
of particles $n_p$ is related to the number density of hydrogen nuclei
$n_{\rm H}$ by $n_p=2(1+\xi)n_{\rm H}$, where
$\xi=0.25Y_{\rm He}/ (1-Y_{\rm He})$ is the ratio of helium nuclei to
hydrogen nuclei, and the baryonic mass fraction of helium is
$Y_{\rm He}=0.24$  (Izotov \& Thuan 1998).
(We have assumed that virtually all the helium is in the form of \HeII.)
For a gas temperature $T$, the thermal energy density of the IGM is then
$\epsilon_T=3(1+\xi)n_{\rm H}kT$, where $k$ is the Boltzmann constant.
After \HeII ionization, $n_p=(2+3\xi)n_{\rm H}$, and the thermal energy density
will be $\epsilon_T'=(3/2)(2+3\xi)n_{\rm H}kT'$, where $T'$ is the gas
temperature immediately following reionization. For a \HeII heating rate
$G_{\rm HeII}$ per \HeII ion and photoionization rate $\Gamma_{\rm HeII}$,
the difference in thermal energies is
$n_{\rm H}\xi G_{\rm HeII}/\Gamma_{\rm HeII}=\epsilon_T'-\epsilon_T$ for
(nearly) complete reionization. The increase in temperature
is then
\begin{equation}
\Delta T=T'-T\simeq\frac{1}{3k}\xi\frac{G_{\rm HeII}}{\Gamma_{\rm HeII}}=
\xi\eta_{\rm HeII}\frac{{\cal E}_{\rm HeII}}{3k},
\label{eq:reheat}
\end{equation}
assuming $2\Delta T>>\xi T'$ and $\xi<<1$. The heating per ionization
efficiency $\eta_{\rm HeII}\equiv G_{\rm HeII}/(\Gamma_{\rm HeII}
{\cal E}_{\rm HeII})$, where ${\cal E}_{\rm HeII}=54.4\,{\rm eV}$ is the
ionization potential of \HeII, is sensitive to the spectral shape of the
ionizing radiation field, which will harden within the ionization front, since
higher energy photons will travel further than lower energy photons before
being absorbed. This results in a boost in $\eta_{\rm HeII}$, and an increase
in $\Delta T$. We find for a radiation field intensity
$J_\nu\propto\nu^{-\alpha}$ within the ionization front that
$0.35<\eta_{\rm HeII}<1.1$ for $2>\alpha>0$. Assuming a hard field with
$\alpha\approx0$ gives $\eta_{\rm HeII}\approx1$ and
$\Delta T\approx1.7\times10^4\kel$.

We estimate the effects of an increase in temperature by broadening the optical
depths in the \LCDML simulation by convolving with a gaussian of effective
velocity dispersion $\sigma_T$. (It may be argued that since the simulation
already incorporates \HeII heating, a smaller temperature increase should be
adopted. On the other hand, the value for $\Omega_bh^2$ is smaller than the
estimate of O'Meara \etal (2000), so that the gas temperature of the moderate
density gas is too low, which partially offsets the additional heating.
In any case, the estimate is only approximate, since it does not account for
the density dependence of the post-photoionization temperature or
any dynamical changes that will result from the
associated sudden change in thermal pressure.) The smoothing formally broadens
the Doppler widths according to $b'=(b^2+2\sigma_T^2)^{1/2}$, where $b$ and
$b'$ are the Doppler parameters prior to and just after the temperature
increase $\Delta T$, respectively. We find that $\sigma_T=12$\kms,
corresponding to $\Delta T=1.7\times10^4\kel$, provides adequate broadening.
The broadening results in a slight increase in $\bar\tau_\alpha$, requiring an
increase in $\Gamma$ to $5\times10^{-13}\, {\rm s}^{-1}$ (or $8\times10^{-13}\,
{\rm s}^{-1}$ re-normalizing to $\Omega_bh^2=0.020$ using $b_{\rm ion}$) at
$z=3.5$ to match the measured value of $\bar\tau_\alpha$,
in good agreement with the estimate of Haardt \& Madau (1996). We
allow for this increase (corresponding to a reduction in $s$ in
Table~\ref{tab:BT97_rescale} from 2.56 to 2.1) when re-synthesizing the model
spectra. We show in Fig.~\ref{fig:BT97_broadened} the effects of the broadening
on the flux per pixel distribution, wavelet coefficients, and absorption line
parameters after synthesizing spectra to match that of Q1937$-$1009, using the
same procedure as previously. The KS
test comparisons with Q1937$-$1009 show excellent agreement with the
distributions of flux per pixel ($\dks=0.018$, $\Pks=0.003$ assuming fully
independent pixels, and $\Pks=0.23$ assuming a reduction by a factor of 3 in
their effective number), the wavelet coefficients on the scale of 17--34\kms
($\dks=0.023$, $\Pks=0.50$) and higher, and the $\NHI$ distribution
($\dks=0.044$, $\Pks=0.28$). Poor agreement is found for the lower velocity
wavelet coefficients and the $b$ distribution because of excessive smoothing
on small velocity scales. The predicted median $b$-value ($\bmed=27.6$\kms),
however, now matches the measured.

We remark that the additional smoothing has only a small effect on the pixel
flux distribution, bringing it into slightly better agreement with the
measured. Much larger effects are produced on the light fluctuations, as
quantified by the wavelet coefficients and absorption line parameters. While
previously the predicted $\NHI$ distribution for the \LCDML was strongly
rejected by the KS test for Q1937$-$1009, it now provides an excellent match.
The predicted number of absorption lines (453) now also agrees with the
measured number (495). The decrease in number is a direct consequence of the
broadening of the lines, as is the good match in the wavelet coefficients
and the improved match in the Doppler parameters. The shape of the $\NHI$
distribution, the number of absorption lines, the wavelet coefficients and the
$b$-values are all linked, changing together in response to the broadening
of the absorption features. We see, then, that it is because the CHDM model
produces broader lines than the other models that it (usually) provides the
best match to these distributions. We caution that conclusions drawn on the
basis of any one of these distributions alone are thus precarious:\ if one of
the distributions is poorly matched, whatever fixes it will alter the others.
By contrast, the flux per pixel distribution is much less sensitive to changes
in the widths of the absorption features, and thus provides a much more robust
test of the cosmological model.

An alternative explanation for the broader measured lines compared with the
model predictions may be galactic feedback. One possibility is Compton heating
by Active Galactic Nuclei (Madau \& Efstathiou 2000). Although Bryan \&
Machacek (2000) found the heating rate inadequate, this may only be
a reflection of the particular parameters adopted for the rate and its
evolution. Another possibility is heating or turbulent broadening by galactic
winds. The possible role of galactic winds in enriching the IGM
with metals has been considered by several authors (Wang 1995;
Nath \& Trentham 1997; Ferrara, Pettini \& Shchekinov 2000;
Madau, Ferrara \& Rees 2000; Aguirre \etal 2001; Theuns, Mo \& Schaye 2001). As
part of the metal-enrichment process, winds from forming galaxies will inject
energy into their surroundings. Although the temperature of the remnant wind
material will decline with time after the winds cease (Madau \etal 2000), the
galactic winds associated with ongoing star-formation will provide a
continuous source of energy to the IGM, broadening the absorption
lines either thermally or through turbulence driven by dynamical
instabilities. Whether or not this is possible without severely disturbing the
agreement with the $\NHI$ distributions is unclear, but we may consider the
energetics required to alter the Doppler parameter distribution. We estimate
the energy density of the \Lya forest as follows. The combined thermal and
kinetic energy per particle in gas giving rise to an absorption
feature with a Doppler parameter $b$ is on the order of
$\epsilon_p\approx(3/4)m_{\rm H}b^2$, where $m_{\rm H}$ is the mass of
a hydrogen atom. Half the baryons are contained in \Lya absorption systems
with $13<\log_{10}\NHI<15$ at $2<z<4$ (Miralda-Escud\'e \etal 1996;
Zhang \etal 1998). For an {\it rms} Doppler parameter of $b\approx25$\kms, this
gives a total (proper) energy density in the \Lya forest at $z=3$ of
$\epsilon_{\rm Ly \alpha}\approx5\times10^{-17}\, {\rm erg\,
cm^{-3}}$. An increase in the Doppler parameter by 20 per cent would require
an increase in the energy density by 40 per cent.
We compare this with the possible
amount of energy injected by supernovae. For a (proper) cosmic star
formation rate density at $z>3$ of $1-10\, {\rm M_\odot\, yr^{-1}\, Mpc^{-3}}$
(Madau \etal 1996; Steidel \etal 1999; Rhoads \etal 2001) that has been
ongoing for $\sim10^9\, {\rm yr}$, a supernova rate of 1 per
$100\, {\rm M_\odot}$ of stars formed, with a kinetic energy of
$10^{51}\,{\rm erg}$ per supernova, would be sufficient to produce a kinetic
energy density input from star formation of
$4\times10^{-17}\, {\rm erg\,cm^{-3}}$, allowing for an
efficiency of 1--10 per cent for the transferal of the kinetic energy
of the supernovae to the IGM. So it is not implausible for feedback from
galaxies in the form of supernova--driven winds to stir the IGM
sufficiently to produce a significant increase in the widths of the
absorption lines. It is possible the presence of the narrow ($b<15$\kms)
absorption systems detected in several of the QSO spectra, of which some if
not all are due to metals, is evidence for supernovae feedback in the form of
enriched material.

\subsection{Gunn-Peterson effect}

The original goal of searches for \Lya absorption in the spectra of QSOs was
the discovery of intergalactic hydrogen, presumed to be uniformly distributed
(Gunn \& Peterson 1965).
The discovery of a uniform component to the IGM has proven largely elusive,
although searches continue. These are pursued under a variety of strategies:\
the search for additional \Lya absorption to what may be accounted for by \Lya
forest blanketing alone (Steidel \& Sargent 1987a), absorption in the
spectra between \Lya forest absorption lines (Giallongo, Cristiani, \& Trevese
1992), or determinations based on the distribution of pixel flux
(Webb \etal 1992; Fang \& Crotts 1995). The measurements provide additional
constraints on the models presented here.
The results of these measurements have generally placed an upper limit on the
optical depth of a smooth component of 0.05--0.1, with a few claimed (but
model-dependent) detections of this order, over the redshift range $2<z<5$.

\begin{figure}
\begin{center}
\leavevmode \epsfxsize=3.3in \epsfbox{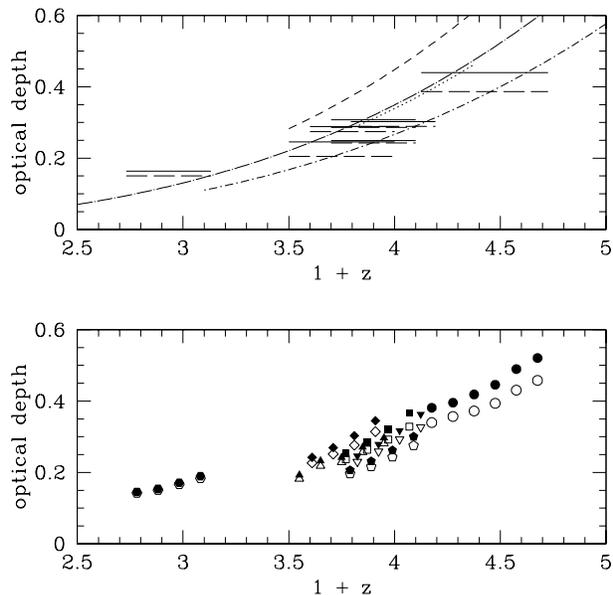}
\end{center}
\caption{(Upper panel)\ Comparison between the total mean optical depth
$\bar\tau_\alpha$ (solid bars) and the effective optical depth $\tau_{\rm eff}$
(long dashed bars) due to line-blanketing by the \Lya forest measured in
the QSO spectra in Table~\ref{tab:qsos}. (The width of a bar indicates
the redshift interval over which the optical depth is averaged.)
Also shown are estimates for
$\bar\tau_\alpha$ using data from Steidel \& Sargent (1987b) (dotted line) and
Zuo \& Lu (1993) (dot-dashed line), the estimate of Press \etal (1993)
(short dashed line), and the estimate of Kim \etal (2001) (dot-long dashed
line). (Lower panel)\ The predictions of the \LCDML model for $\bar\tau_\alpha$
(solid symbols) and $\tau_{\rm eff}$ (open symbols) for the spectra. The points
near $z=3$ have been slightly offset in redshift for clarity.
The symbols correspond to Q1937$-$1009 (circles), HS~1946$+$7658 (triangles),
Q0014$+$813 (inverted triangles), Q0302$-$003 (squares), Q0636$+$680
(diamonds), Q0956$+$122 (pentagons), J2233$-$606 (hexagons).}
\label{fig:taudiff}
\end{figure}

The models discussed here show that essentially all the structure of
the IGM is in the form of fluctuations that give rise to absorption
features, even in underdense regions (Zhang \etal 1998), although a
residual smooth component may be detectable in minivoids. A small amount
of residual absorption may be reconciled with BBN constraints on $\Omega_b$
and a photoionizing background dominated by QSO sources, provided the systems
giving rise to the \Lya forest have a characteristic line-of-sight
scale length of $c_s/H(z)\approx L_J$, so that most of the baryons
are contained in the \Lya
forest (Meiksin \& Madau 1993). This is explicitly demonstrated by the
simulations. We may estimate the amount of additional absorption due to a
smooth component as the difference $\tau_s=\bar\tau_\alpha-\tau_{\rm eff}$,
where $\tau_{\rm eff}$ is the effective optical depth due to
line-blanketing, given over a redshift interval $\Delta z$ by
$\tau_{\rm eff}=(1/\lambda_\alpha)\sum_i w^{\rm obs}_i/ \Delta z$,
where the sum is carried over the observed equivalent widths of the
lines and $\lambda_\alpha$ is the (rest) wavelength of \Lya
(Meiksin \& Madau 1993; Press, Rybicki \& Schneider 1993). 
The mean and effective optical depths measured
from the spectra in Table~\ref{tab:qsos} (with metal lines removed as
discussed in Section~\ref{sec:results}) are shown in Figure~\ref{fig:taudiff}.
The differences in optical depth range over
$0.01<\tau_s<0.05$. The predictions of the \LCDML model are also shown.
We find $\tau_s<0.1$ at $z\simeq3.5$, diminishing nearly to zero at lower
redshift. Thus there is very little absorption present in both the measured
and the model spectra that may not be accounted for by the \Lya forest alone.
Similar results are found for the other models.

We also show in Figure~\ref{fig:taudiff} estimates for $\bar\tau_\alpha$ from
Press, Rybicki \& Schneider (1993) and Zhang \etal (1997), who based their
estimates on the mean absorption measurements of Steidel \& Sargent (1987b) and
Zuo \& Lu (1993). These measurements have been used in the past for estimating
the amount of any residual optical depth after removing the effect of
line-blanketing. The estimates of Press \etal (1993) lie systematically high,
and those of Zuo \& Lu (1993) lie systematically low, compared with the values
we derive. Our values agree closely with those derived from the measurements
of Steidel \& Sargent (1987b), and with the recent determination by
Kim, Cristiani \& D'Odorico (2001). Adopting the values of Press \etal (1993)
would in particular result in an overestimate of any residual optical depth
beyond that due to line-blanketing.

\section{Summary}
\label{sec:summary}

We compare measurements of the \Lya forest in seven high resolution,
high signal-to-noise ratio QSO spectra with predictions of
hydrodynamical simulations for a variety of CDM dominated
cosmologies. Spectra are synthesized from the simulations matching the
observed spectra in wavelength coverage, pixelization, spectral resolution,
and signal-to-noise properties. The statistical comparisons between the model
and measured spectra are based on the distribution of flux
per pixel and the statistics of the light fluctuations in the spectra
as quantified by a wavelet analysis and by Voigt profile fitting to
absorption features. Although all the models provide reasonably good
qualitative descriptions of the measured \Lya forest properties, we
find varying levels of agreement with the data depending both on the
cosmological model and the basis of comparison. In general, no single
model is able to pass all the tests. The most accurate agreement is obtained
for the flux per pixel distributions, with differences between the predicted
and measured cumulative flux distributions as small as $\dks<0.02$, close to
the precision limit of the simulations and the measurements. Although the KS
test formally rejects (at the $3\sigma$ level) almost all the predicted flux
distributions in comparison with the measured, an analysis of the probability
distribution of $\dks$ shows evidence for significant correlations in the flux
values of neighbouring pixels. Allowing for the indicated reduction in the
effective number of independent pixels results in acceptable fits provided
by the \LCDM and SCDM models, while the remaining models are strongly rejected.
We find that the slope of the flux per pixel distribution systematically
decreases with increasing $\sigma_{\rm J}$, the density fluctuations on the
scale of the cosmological Jeans length. Agreement between the predicted and
measured flux per pixel distributions imposes the restriction
$1.3<\sigma_{\rm J}<1.7$ (better than $3\sigma$), normalized at $z=3$.

The predicted light fluctuations, as quantified by the wavelet analysis and
fit absorption line parameters, are generally in poor agreement with the data.
The most stringent statistical constraints are set by the distributions of
wavelet coefficients, because of the large number of pixels available.
(Correlations between the wavelet coefficients are weak or absent.)
An advantage of the
wavelet analysis over the more traditional Voigt profile analysis is that the
orthogonality of the wavelets permits a separation of the velocity 
scales contaminated by metal lines from those dominated by \Lya absorption
features. A further advantage is that it is extremely computationally cheap
relative to a Voigt profile analysis, providing a very inexpensive means of
testing model predictions for the statistics of the light fluctuations to
high precision. The velocity range of approximately 15--30\kms is of special
interest since it corresponds to the velocity broadening of hydrogen
absorption lines at the characteristic temperature of photoionized gas.
Only the CHDM model is able to provide an acceptable
match to the measured coefficients on this scale; all other models are strongly
rejected. A conflict is thus presented in relation to the flux distribution:\
models that match well the measured flux per pixel distributions provide
unacceptable matches to the wavelet coefficients, while those that are in
agreement with the measured wavelet coefficients produce unacceptable pixel
flux distributions. The agreement at smaller velocity scales is mixed, possibly
due to contamination by metal absorption lines.

Voigt profile analyses are also performed, as they relate the light
fluctuations to quantities of direct physical interest, the column densities
and velocity widths of the absorption features. We find that while good
agreement is often found between the predicted and measured $\NHI$
distributions, including for models that produce satisfactory flux per pixel
distributions, the agreement is mixed, depending on the analyis algorithm and
the data. When agreement is found, it is often marginal. The agreement
generally tends to favour the CHDM model, as for the wavelet coefficients. The
same is true for the total number of absorption features. The models other than
CHDM usually predict far too many absorption lines compared with the number
measured in the spectra. The model predictions for the Doppler parameters fare
much worse. The predicted median Doppler parameters are as much as 60 per cent
smaller than the measured. As for the wavelet coefficients, the
best agreement is generally provided by the CHDM model, although the predicted
distributions are generally statistically unacceptable.

We find considerable inhomogeneity in the published data, particularly on the
15--30\kms scale. Doppler parameter distributions independently derived from
separate spectra covering nearly the same redshift range show statistically
significant differences. This may lead to an exaggeration in the discrepancy
between the Doppler widths found in simulations and the measured values
appearing in some of the literature.

A breakdown of the absorption lines into subsamples optically thin and
thick at the \Lya line centre shows that the discrepancies between the
predicted and measured absorption line parameters originate
primarily from the optically thin absorbers. The $\NHI$ distributions
predicted for the optically thick systems by most of the models 
are generally in good agreement with the measured distributions,
as are the numbers of optically thick lines. Much poorer
agreement is found for the $\NHI$ distributions and line numbers for the
optically thin absorbers. The Doppler parameters show a similar trend. The
$b$-values of the optically thick absorbers are in much better accord with the
data than those of the optically thin systems. The CHDM model again tends
to provide the best agreement for both the $b$ distributions and the total
number of optically thin systems, as it did for the wavelet coefficients on the
16--32\kms scale. This is no surprise, as all these distributions are related.
The CHDM model produces broader features than the other models, so that fewer
lines are required to reproduce the measured mean optical depth. The fewer
lines are reflected by proportionally fewer large
wavelet coefficients (fewer large light fluctuations), agreeing more
closely with the measured coefficient distributions.

We consider a few possible explanations for the discrepant results for
the light fluctuations. The most likely is perhaps late reionization
of \HeII, at $z_{\rm HeII}\approx3.5$. Late reionization will increase the
temperature of the gas by an amount $\Delta T\approx1.5-2\times10^4\kel$. While
the temperature of overdense gas will relax to lower values as a result of
radiative losses, the lower density gas giving rise to the optically thin
systems will retain a stronger memory of the boost in temperature, because it
is too rarefied to radiate efficiently. We estimate the impact of the
temperature boost for the \LCDML model predictions for Q1937$-$1009 by
convolving the simulation optical depths with a gaussian of velocity
width $\sigma_T=12$\kms, corresponding to $\Delta T=1.7\times10^4\kel$.
Excellent agreement is produced for the distributions of pixel flux,
wavelet coefficients (on the 17--34\kms scale), \HI column density, and the
median Doppler parameter. The extra broadening of the features has very little
effect on the flux distribution, suggesting that the flux per pixel
distribution provides a robust test of the models that is insensitive to
uncertainties in the modelling of the reionization history of the IGM.
We also consider the possible impact of galactic winds on the IGM
and show that they too may plausibly provide sufficient broadening of the
absorption features, provided they are able to propagate throughout the low
density regions which give rise to the optically thin systems.

We caution against the use of any single distribution quantifying the
light fluctuations, such as the wavelet coefficients, the \HI column density,
the Doppler parameters, or the number of absorption lines, as a test of the
models. These quantities are all related, so that if a model fails to
reproduce any one of them, whatever alterations to the model are necessary to
gain accord with the data will disturb any agreement found in the remaining
quantities.

The requirement that the models match the measured mean optical depths
of the \Lya forest places a constraint on the \HI photoionization
rate. The models presented require an ionization rate that lies within
a factor of 2 of that predicted for a UV background dominated by QSO sources.
A large contribution from additional sources of ionizing
radiation would place the models in jeopardy by requiring
an implausibly high baryon density. We note, however, that the
required photoionization background declines at $z>3.1$ significantly
more slowly than predicted by Haardt \& Madau (1996) for a background
dominated by QSO sources, based on current QSO counts. The mean
optical depths obtained by the models are dominated by line-blanketing, with a
residual optical depth resulting from a uniform component of at most
0.1 at $z\simeq3.5$, and becoming vanishingly small by $z\simeq2$.

\section*{Acknowledgments}

This work is supported in part by the NSF under the auspices of the
Grand Challenge Cosmology Consortium (GC$^3$). The computations
were performed on the Convex C3880, the SGI Power Challenge,
and the Thinking Machines CM5 at the National Center for Supercomputing
Applications, and the Cray C90 at the Pittsburgh Supercomputing Center
under grant AST950004P. AM thanks S. Burles and D. Tytler for providing an
unpublished electronic version of the spectrum of Q1937$-$1009, and R. Dav\'e
for permission to use \AutoVP. Support for GLB was provided by NASA
through Hubble Fellowship grant HF-01104.01-98A from the Space Telescope
Science Institute, which is operated by the Association of Universities for
Research in Astronomy, Inc., under NASA contract NAS6-26555. MM would like
to acknowledge the hospitality of the MIT astrophysics group.

\begin{table}
\centerline{\begin{tabular}{|c|c|c|c|c|c|c|c|} \hline
Model & $\Omega_M$ & $\Omega_\Lambda$   & $\Omega_bh^2$ & $h$   & $n$        & $\sigma_{8h^{-1}}$ & $\sigma_{\rm J}$\\
\hline 
\hline
CHDM   & 1   & 0   & 0.025   & 0.6  & 0.98 & 0.7 & 1.1 \\ \hline
\LCDML & 0.4 & 0.6 & 0.015   & 0.65 & 1    & 1.0 & 1.7 \\ \hline
\LCDMH & 0.4 & 0.6 & 0.021   & 0.65 & 1    & 0.8 & 1.3 \\ \hline
OCDM   & 0.4 & 0   & 0.015   & 0.65 & 1    & 1.0 & 2.2 \\ \hline
SCDM   & 1   & 0   & 0.015   & 0.5  & 1    & 0.7 & 1.6 \\ \hline
tCDM   & 1   & 0   & 0.025   & 0.6  & 0.81 & 0.5 & 0.9 \\ \hline
\end{tabular}}
\caption{Parameters for the cosmological models.
$\Omega_M$ is the total mass density parameter, $\Omega_\Lambda$ the 
cosmological constant density parameter, $\Omega_b$ the baryonic mass fraction,
$h=H_0/ 100$\kmsmpc, where $H_0$ is the Hubble constant at $z=0$,
$n$ the slope of the primordial density perturbation power spectrum,
$\sigma_{8h^{-1}}$ the density fluctuation normalization in a sphere of
radius $8h^{-1}$ Mpc, and $\sigma_{\rm J}$ the density fluctuation at $z=3$
on the scale of the cosmological Jeans length (see text). For the CHDM model,
two massive neutrino species are assumed accounting for a mass fraction
of $\Omega_\nu=0.2$.
}
\label{tab:model_par}
\end{table}

\begin{table}
\centerline{\begin{tabular}{|l|c|c|c|c|} \hline
QSO & $z_{\rm em}$ & $z$ range & pixel res & spectral res\\
\hline
\hline
Q1937$-$1009 & 3.806 & $3.126-3.726$ & $7.4\times10^4$ & 9.0 \\
HS~1946$+$7658 & 3.051 & $2.500-3.000$ & $1.5\times10^5$ & 7.9 \\
Q0014$+$813  & 3.384 & $2.794-3.194$ & $0.06\,{\rm \AA}$ & 8.3 \\
Q0302$-$003  & 3.286 & $2.700-3.100$ & $0.06\,{\rm \AA}$ & 8.3 \\
Q0636$+$680  & 3.174 & $2.600-3.000$ & $0.06\,{\rm \AA}$ & 8.3 \\
Q0956$+$122  & 3.301 & $2.700-3.100$ & $0.06\,{\rm \AA}$ & 8.3 \\
J2233$-$606  & 2.238 & $1.732-2.132$ & $0.05\,{\rm \AA}$ & 6.7 \\ \hline
\end{tabular}}
\caption{The QSO sample. The indicated redshift range is the portion of the
spectrum adopted for the analysis. The pixel resolution (as
$\lambda/\Delta\lambda$ or pixel width) and spectral resolution (FWHM in \kms),
correspond to this region. The observations were reported
in:\ Q1937$-$1009:\ Burles \& Tytler (1997);
HS~1946$+$7658:\ Kirkman \& Tytler (1997); Q0014$+$813,
Q0302$-$003, Q0636$+$680, Q0956$+$122:\ Hu, Kim, Cowie \& Songaila (1995);
J2233$-$606:\ Cristiani \& D'Odorico (2000).
}
\label{tab:qsos}
\end{table}

\begin{table}
\centerline{\begin{tabular}{|c|c|c|c|c|} \hline
Model & $\alpha$ & $p$ & $\omega$ & $s$ \\
\hline
\hline
CHDM & 6.3 & 0.0 & 0     & 0.3274 \\
CHDM & 6.3 & 5.0 & 0     & 0.5525 \\
CHDM & 6.3 & 5.0 & 0.04  & 0.63 \\ \hline
$\Lambda$CDM$_{\rm L}$ & 7.4 & 0.0 & 0     & 1.465 \\
$\Lambda$CDM$_{\rm L}$ & 7.4 & 5.0 & 0     & 2.443 \\
$\Lambda$CDM$_{\rm L}$ & 7.4 & 5.0 & 0.013 & 2.56 \\ \hline
$\Lambda$CDM$_{\rm H}$ & 6.8 & 0.0 & 0     & 0.5218 \\
$\Lambda$CDM$_{\rm H}$ & 6.8 & 5.0 & 0     & 0.8783 \\
$\Lambda$CDM$_{\rm H}$ & 6.8 & 5.0 & 0.022 & 0.94 \\ \hline
OCDM & 7.4 & 0.0 & 0     & 3.119 \\
OCDM & 7.4 & 4.5 & 0     & 4.937 \\
OCDM & 7.4 & 4.5 & 0.005 & 4.937 \\ \hline
SCDM & 7.0 & 0.0 & 0     & 1.291 \\
SCDM & 7.0 & 5.0 & 0     & 2.105 \\
SCDM & 7.0 & 5.0 & 0.018 & 2.3 \\ \hline
tCDM & 6.4 & 0.0 & 0     & 0.3076 \\
tCDM & 6.4 & 5.2 & 0     & 0.5260 \\
tCDM & 6.4 & 5.2 & 0.056 & 0.63 \\
\hline
\end{tabular}}
\caption{Evolution exponents, continuum offsets, and rescaling coefficients for
the predicted distributions of flux per pixel to match that measured
in Q1937$-$1009. The value for $\alpha$ corresponds to the redshift interval
$3<z<4$.}
\label{tab:BT97_rescale}
\end{table}

\begin{table}
\centerline{\begin{tabular}{|c|c|c|c|} \hline
Model & $\alpha$ & $\omega$ & $s$ \\
\hline
\hline
CHDM & 6.3 & 0.03  & 0.59 \\
$\Lambda$CDM$_{\rm L}$ & 6.8 & 0.005 & 2.1 \\
$\Lambda$CDM$_{\rm H}$ & 6.8 & 0.015 & 0.85 \\
tCDM & 6.4 & 0.04  & 0.59 \\
\hline
\end{tabular}}
\caption{Evolution exponents, continuum offsets, and rescaling coefficients
for the predicted distributions of flux per pixel to match that
measured in HS~1946$+$7658. The
value for $\alpha$ corresponds to the redshift interval $2<z<3$.}
\label{tab:KT97_rescale}
\end{table}

\begin{table}
\centerline{\begin{tabular}{|c|c|c|} \hline
Model & $\omega$ & $s$ \\
\hline
\hline
& Q0014$+$813 \\
CHDM & 0.08  & 0.6263 \\
$\Lambda$CDM$_{\rm L}$ & 0.03 & 2.2 \\
$\Lambda$CDM$_{\rm H}$ & 0.05 & 0.90 \\
tCDM & 0.08  & 0.60 \\
\hline
& Q0302$-$003 \\
CHDM & 0.075  & 0.74 \\
$\Lambda$CDM$_{\rm L}$ & 0.03 & 2.7 \\
$\Lambda$CDM$_{\rm H}$ & 0.05 & 1.09 \\
tCDM & 0.08  & 0.72 \\
\hline
& Q0636$+$680 \\
CHDM & 0.075  & 0.82 \\
$\Lambda$CDM$_{\rm L}$ & 0.026 & 2.90 \\
$\Lambda$CDM$_{\rm H}$ & 0.045 & 1.15 \\
tCDM & 0.08  & 0.80 \\
\hline
& Q0956$+$122 \\
CHDM & 0.06  & 0.54 \\
$\Lambda$CDM$_{\rm L}$ & 0.025 & 1.88 \\
$\Lambda$CDM$_{\rm H}$ & 0.045 & 0.81 \\
tCDM & 0.08  & 0.57 \\
\hline
& J2233$-$606 \\
CHDM & 0.06  & 1.3 \\
$\Lambda$CDM$_{\rm L}$ & 0.024 & 4.5 \\
$\Lambda$CDM$_{\rm H}$ & 0.04 & 1.95 \\
tCDM & 0.06  & 1.25 \\
\hline
\end{tabular}}
\caption{Continuum offsets and rescaling coefficients for the predicted
distributions of flux per pixel to match those measured in
Q0014$+$813, Q0302$-$003, Q0636$+$680, Q0956$+$122, and J2233$-$606.}
\label{tab:Hu95_rescale}
\end{table}

\appendix
\section{Tests of analysis procedures}
\label{app:tests}

In this section we summarise several tests of the analysis procedures:\ the
redshift interpolation method, the noise models, the goodness-of-fit of the
absorption line analyses, and the applicability of the KS test.

\subsection{Redshift interpolation}

\begin{figure}
\begin{center}
\leavevmode \epsfxsize=3.3in \epsfbox{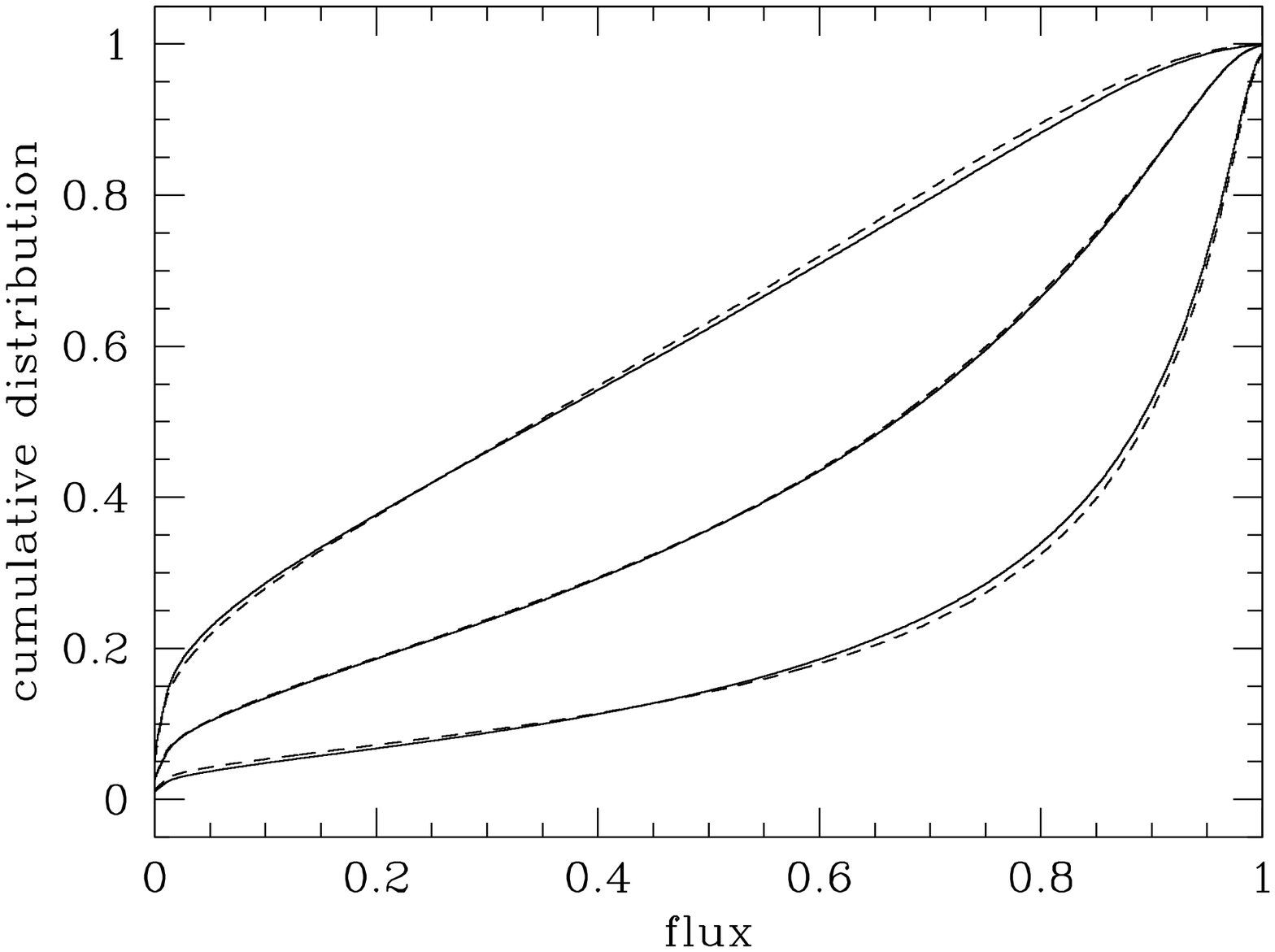}
\end{center}
\caption{Test of the interpolation method for the flux distribution, for
the \LCDMH model. The left pair of curves show the flux distribution at
$z=4$ (solid curve) and the predicted distribution based on rescaling the
model spectra at $z=3$ according to equation~\ref{eq:rescale} (dashed curve).
The right pair of curves show the flux distribution at $z=3$ (solid curve) and
the predicted distribution from rescaling the spectra at $z=4$ (dashed curve).
The central pair show the flux distribution at $z=3.6$ (solid curve) and the
linearly interpolated distribution based on rescaling the spectra at $z=3$ and
$z=4$ and using equation~\ref{eq:interp} (dashed curve). The predicted and
actual distributions at $z=3.6$ agree to within 0.4 per cent.}
\label{fig:interp_test}
\end{figure}

We test the redshift interpolation method using the distributions found in
the \LCDMH simulation. In Figure~\ref{fig:interp_test}, the effect of rescaling
the flux distributions at $z=3$ and $z=4$ to redshifts $z=4$ and $z=3$,
respectively, is shown using the values for $\alpha$ in
Table~\ref{tab:BT97_rescale} and equation~\ref{eq:rescale}. (No additional
evolution in the radiation field was assumed:\ $p=0$.) The result of rescaling
the spectra at $z=3$ and $z=4$ both to $z=3.6$ and then linearly weighting
the resulting flux distributions to $z=3.6$ using equation~\ref{eq:interp} is
shown by the central dashed curve. The difference between the direct and
interpolated cumulative distributions at $z=3.6$ is at most 0.006, and
indicates the level of accuracy of the modelling of the measured spectra. A
similar level of accuracy is found for the cumulative distributions of the
wavelet coefficients. The direct and interpolated distributions of the
absorption line parameters, however, agree somewhat more poorly. While the
cumulative distributions generally agree to within a difference of 0.01,
portions may disagree by as much as 0.03 for the $\NHI$ distribution and 0.04
for the $b$ distribution at intermediate values, in the sense that the
interpolated distributions exceed the directly estimated distributions. These
differences are comparable to the accuracy to which the distributions may be
measured using a single QSO spectrum and represent a limitation of the
modelling of the distributions. Since it was based on a near maximum amount of
interpolation to an intermediate redshift between the data dump redshifts, we
expect that the actual modelling is somewhat better. Nevertheless, in the
future a greater number of intermediate data dump redshifts would be desirable.

\subsection{Absorption line goodness-of-fit}

\begin{figure}
\begin{center}
\leavevmode \epsfxsize=3.3in \epsfbox{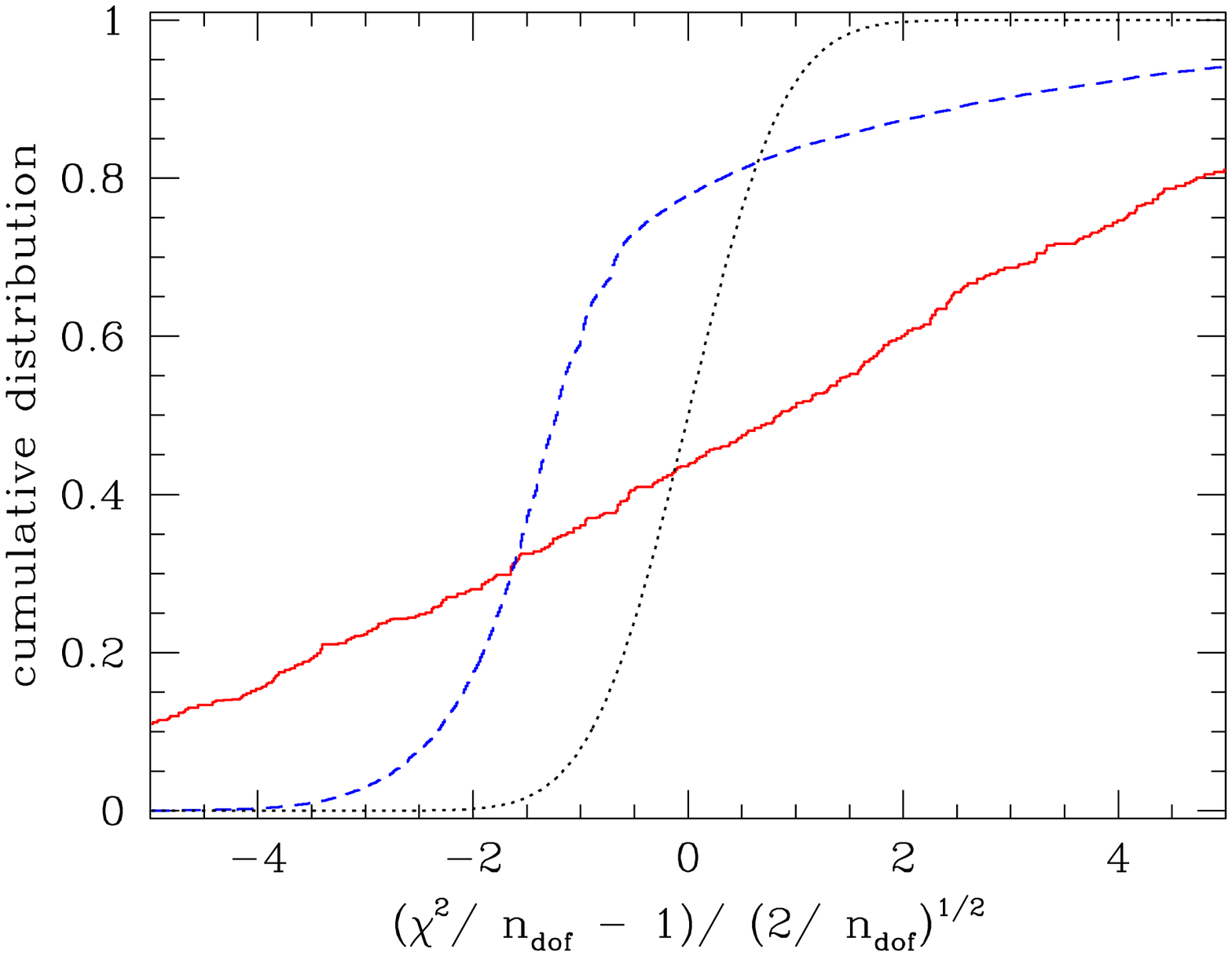}
\end{center}
\caption{Distribution of $\chi^2$ for the absorption line fits using \AutoVPfc\
(solid curve) and \SPECFITfc\ (dashed curve), for \LCDMLns. Shown are the
cumulative distributions of $(\chi^2/ n_{\rm dof}-1)/ (2/ n_{\rm dof})^{1/2}$,
where $n_{\rm dof}$ is the number of degrees-of-freedom of the fit. This
quantity should be distributed like a gaussian with vanishing mean and unit
variance in the limit $n_{\rm dof}\rightarrow\infty$ (dotted curve). Although
most of the fits are acceptable, both methods produce too great a tail at
large $\chi^2$ values.}
\label{fig:chisq_test}
\end{figure}

Because of the necessarily automated nature of the absorption line parameter
analyses, it is important to maintain a check on the acceptability of the fits.
This was done by tabulating the distributions of $\chi^2$ for the fit line
complexes for all of the analyses. Because the number of degrees-of-freedom
$n_{\rm dof}$ differs for different line complexes, the quantity tabulated is
$(\chi^2/ n_{\rm dof}-1)/ (2/ n_{\rm dof})^{1/2}$, which in the limit
$n_{\rm dof}\rightarrow\infty$ is distributed like a gaussian with zero
mean and unit variance (Kendall \& Stuart 1969). The resulting cumulative
distributions for the \LCDML model predictions for Q1937$-$1009 (assuming no
continuum offset) are shown for \AutoVP\ and \SPECFIT\ in
Figure~\ref{fig:chisq_test}, which are representative of all the analyses.
The fits are generally acceptable, although both line analysis methods produce
a tail of high $\chi^2$ values. The low median value of $\chi^2$ produced by
\AutoVP\ is a consequence of the demand for a reduced $\chi^2$ of 0.5 for a fit
to be acceptable. For \SPECFIT, the requirement was a reduced $\chi^2$ of 1.

\subsection{KS test}

\begin{figure}
\begin{center}
\leavevmode \epsfxsize=3.3in \epsfbox{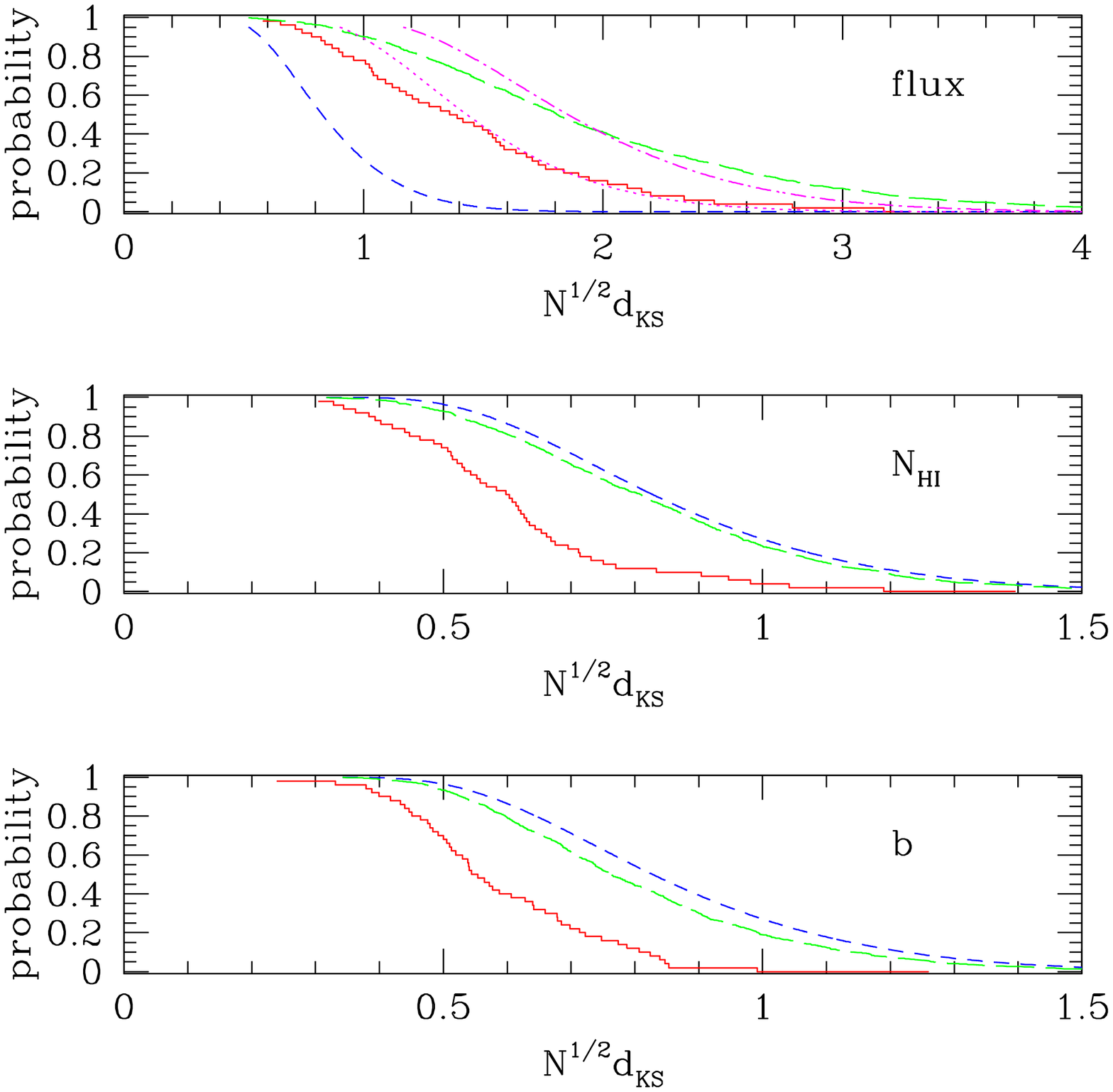}
\end{center}
\caption{Test of the distribution of probability of exceeding a given value of
$N^{1/2}\dks$, where $\dks$ is the maximum difference between the cumulative
distribution of $N$ values of flux (top panel), $\NHI$ (middle panel), and $b$
(bottom panel), and the expected distributions, using the predictions of the
\LCDML model for Q1937$-$1009. The probability distribution derived from the
simulation (solid histograms) is broader for the flux, and narrower for the
absorption line parameters, than the respective KS probability distributions
(short dashed curves). The distributions for a set of Monte Carlo realisations
of the \Lya forest are also shown (long dashed curves). Also shown in the top
panel are the KS probability curves assuming a reduction in the effective
number of independent pixels by a factor of 3 (dotted curve) and 5
(dash-dotted curve).}
\label{fig:KSprob_test}
\end{figure}

The acceptability of a hypothesis using the KS test is determined by
the probability of exceeding the value $N^{1/2}\dks$, where $\dks$ is the
maximum difference (in absolute value) between the cumulative
distribution of some parameter measured independently $N$ times, and
the predicted distribution. The probability distribution for exceeding
$N^{1/2}\dks$ in the flux per pixel distribution
derived from the simulated spectra of the \LCDML model
(with no continuum offset), using as the fiducial the average
distribution of flux per pixel predicted for Q1937$-$1009, is shown in
Figure~\ref{fig:KSprob_test}. The distribution of $N^{1/2}\dks$
derived from the simulated spectra is broader than the theoretical KS
probability distribution. Also shown are the probability distributions
of $N^{1/2}\dks$ based on the distributions of $\NHI$ and $b$ from the
\AutoVP\ analysis. In contrast to the case for the flux, these
distributions are much narrower than the theoretical distribution.

The reasons for the discrepancies in the distributions are unclear. Since
absorption features typically run over several pixels, individual pixel fluxes
will not be statistically independent, as is required for the validity of the
KS test. Similarly, since the absorption line analysis procedure fits nearby
lines simultaneously, the values found for the fit parameters are not truly
independent.

We examine the possible effect of correlations in flux and in the
derived absorption parameter values using a set of Monte Carlo
simulations of spectra.  We model the statistical distribution of the
line positions, \HI column densities, and Doppler parameters based on
the distributions reported in Kirkman \& Tytler (1997) (see
Appendix~\ref{app:lpcomp} below). The resulting probability
distribution of $N^{1/2}\dks$ for the flux, shown in
Figure~\ref{fig:KSprob_test}, is again found to be broader than the KS
distribution, and even broader than the distribution derived from the
simulation. This suggests that strong correlations in the pixel fluxes
are present. Allowing for a reduction by a factor of 3 in the effective
number of independent flux measurements shifts the theoretical KS probability
distribution into good agreement with that derived from the simulation, while
a reduction by a factor of 5 produces good agreement between the theoretical
probability distribution and that measured from the Monte Carlo realizations.
The resulting probability distributions of $N^{1/2}\dks$ based on the
derived values of $\NHI$ and $b$ using \AutoVP\ are found to be narrower
than the KS distribution, although not by as much as the simulation results.
Once again, correlations appear to be present, but tending to reduce the
scatter in the derived values of the absorption line parameters.

An element of statistical dependence in the simulation results may also arise
because the lines-of-sight sampled in the simulation are not truly
statistically independent since they are drawn from the same simulation, and
so may give a biased estimate of the correct distribution function with a
reduced dispersion. This may only be tested using lines-of-sight that are
spaced adequately to probe separate structures, or by a large number of
repeated simulations. Neither is currently computationally feasible using full
hydrodynamical simulations, so that this possibility cannot be tested at
present.

Still another possibility is that the properties of nearby features are
physically related. Such correlations are expected for neighbouring
lines--of--sight as a consequence of the linear continuity of the gas density
defining the filaments. Similar correlations may exist among density
fluctuations across the filaments. In this paper, we quote formal KS
probabilities, noting that these appear to underestimate the true probabilities
for the flux distribution comparisons and to be conservative estimates of the
probabilities for the comparisons of the absorption line parameter values.
Possibly more realistic estimates of the flux per pixel distribution KS-test
probabilities are provided by reducing the effective number of pixels by a
factor of 3--5, and so these values are also quoted.

\section{\bf Comparison of \AutoVPbf\ {\bf and} \SPECFITbf}
\label{app:lpcomp}

In this Appendix, the line finding and fitting algorithms used in this
paper, \AutoVP\ and \SPECFIT, are compared using Monte Carlo
realisations of spectra. The algorithm for \AutoVP\ is described by
Dav\'e \etal (1997). \SPECFIT\ performs its analysis through the following
steps:\ 1.\ filter the spectrum using wavelets, discarding the
smallest wavelet coefficients to ensure a reduced $\chi^2$ of 1
between the filtered and unfiltered spectrum; 2.\ identify candidate
features as inflection points by computing a smoothed second
derivative of the spectrum; 3.\ define a spectral region to be fit
about each candidate line as a contiguous region with the flux smaller
than a given $\exp(-\tau_{\rm min})$; 4.\ merge overlapping regions
into single regions, provided the number of candidates in a single
region is not too great, in which case it is split (at most 16
candidate lines were allowed in a single fit region in the analyses of
this paper); 5.\ perform a non-linear least squares fit of the
candidate lines to the original (unfiltered) spectrum.

The spectra are constructed from discrete lines with Voigt profiles
using the \HI column density and Doppler parameter distributions found
by Kirkman \& Tytler (1997). Specifically, the \HI column densities $\NHI$
are drawn independently from a power law distribution of slope 1.5 between
$12.5<\log_{10}N_{\rm HI}<16$ and the Doppler parameters $b$ from a
gaussian with mean 23~\kms and standard deviation 14~\kms. A cut--off
in $b$ is imposed according to $b>14 + 4(\log_{10} N_{\rm HI} -
12.5)$~\kms. The resulting average Doppler parameter is 31~\kms. The
number density of lines per unit redshift matches that of Kirkman \&
Tytler (1997) at $z=3$. The resolution is set at
$\lambda/d\lambda=5\times10^4$, and gaussian noise is added to give
a continuum signal--to--noise ratio per pixel of 50.

\begin{figure}
\begin{center}
\leavevmode \epsfxsize=3.3in \epsfbox{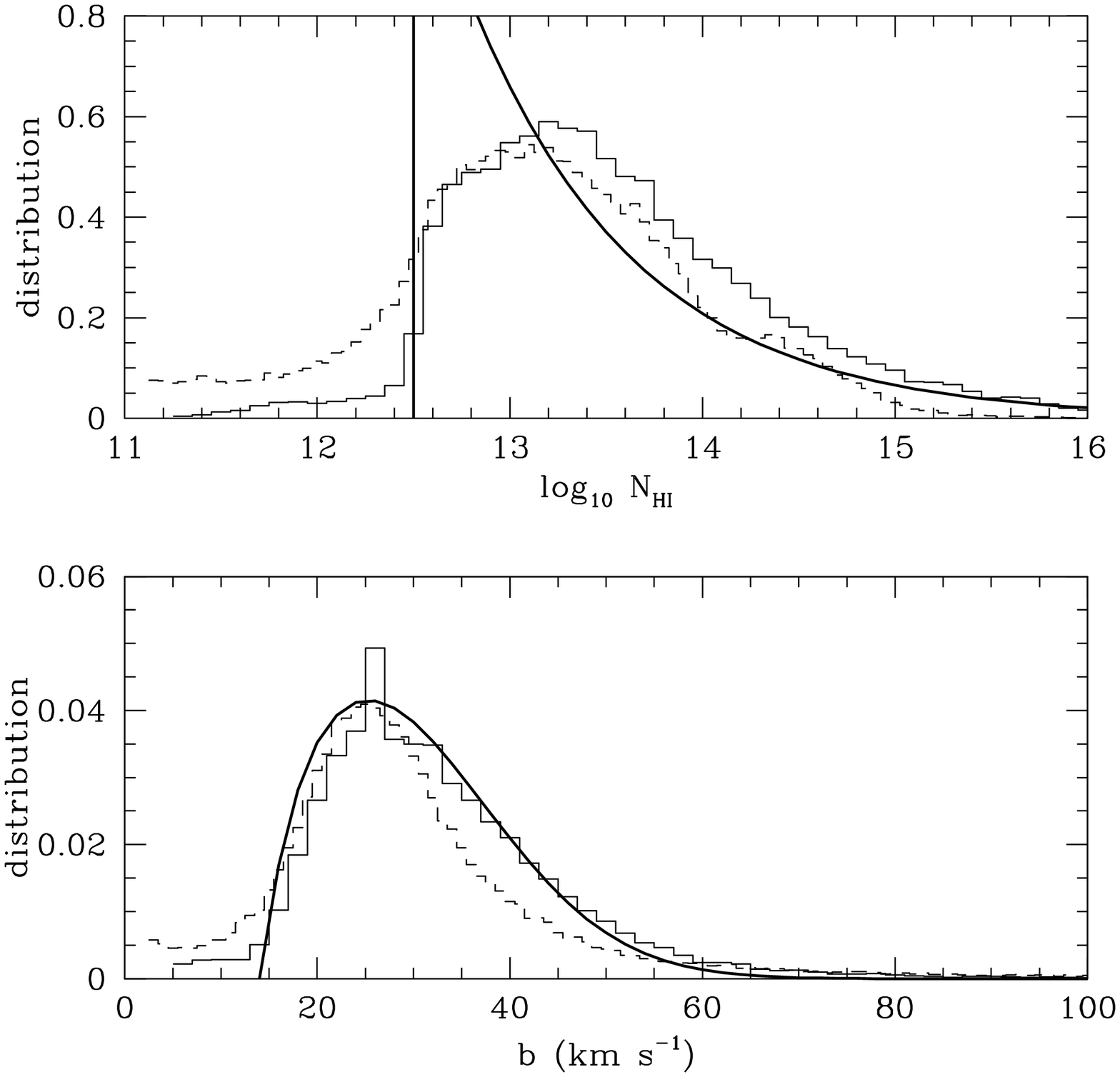}
\end{center}
\caption{The recovered \HI column density and Doppler parameter distributions
from a set of Monte Carlo realizations. The light solid curves show the results
of \AutoVPfc\ while the dashed curves show the results of \SPECFITfc. The heavy
solid lines show the input model distributions.}
\label{fig:avp_spf_comp}
\end{figure}

The two methods are found to have relative advantages and disadvantages.
\AutoVP\ faithfully recovers the cut-off in the \HI column density distribution
at $\log_{10}N_{\rm HI}=12.5$, while \SPECFIT\ produces an excess
of low column density systems. At higher column densities
($13<\log_{10}N_{\rm HI}<15$), however, \SPECFIT\ is found to better recover
the input column density distribution.

By contrast, \AutoVP\ well reproduces the input Doppler parameter distribution,
although it produces a small tail at low values. \SPECFIT\ similarly produces
a low $b$ tail, and somewhat underpredicts the number of intermediate $b$
systems.

It is evident that neither algorithm perfectly recovers the input parameter
distributions. No improvement in the signal--to--noise ratio would yield better
results:\ the discrepancy arrises from the unavoidable blending of features.
In practice, provided the same algorithm is applied to the data and the model
spectra, the resulting fits may still be used as a viable statistical
description of the light fluctuations in the spectra.


\end{document}